\documentclass[aps,prd,twocolumn,superscriptaddress,nofootinbib,reprint]{revtex4-2}
\usepackage[dvipsnames]{xcolor}
\usepackage{graphicx}
\usepackage{amsmath}
\usepackage{amsfonts}
\usepackage{dsfont}
\usepackage[german, english]{babel}
\usepackage{ulem}
\usepackage{color, xcolor}
\usepackage{float}
\usepackage{physics }
\usepackage[colorlinks=true,citecolor=blue,urlcolor=blue,linkcolor=blue]{hyperref}
\usepackage{bbold}
\usepackage{nccmath}
\usepackage{amssymb}

\newcommand{\secref}[1]{Sec.~\ref{#1}}
\newcommand{\appref}[1]{Appendix~\ref{#1}}
\newcommand{\figref}[1]{Fig.~\ref{#1}}
\renewcommand{\eqref}[1]{Eq.~(\ref{#1})}

\newcommand{\bk}{\boldsymbol{k}}
\newcommand{\bx}{\boldsymbol{r}}
\newcommand{\sext}{s_\text{ext}}
\newcommand{\bz}{\text{BZ}}
\newcommand{\da}{^\dagger}
\renewcommand{\max}{\text{max}}

\begin{document}

\title{{Microscopic theory of strain-controlled split superconducting and time-reversal symmetry-breaking transitions in $s+id$ superconductor}}
\author{Anton Talkachov}
\email{anton.talkachov@gmail.com}
\affiliation{Department of Physics, KTH-Royal Institute of Technology, SE-10691, Stockholm, Sweden}
\author{Egor Babaev}
\affiliation{Department of Physics, KTH-Royal Institute of Technology, SE-10691, Stockholm, Sweden}
\affiliation{Wallenberg Initiative Materials Science for Sustainability, Department of Physics, KTH Royal Institute of Technology, SE-106 91 Stockholm, Sweden}
\date{\today}

\begin{abstract}

We study conditions of the appearance of $U(1)\times \mathbb{Z}_2$ superconducting states that spontaneously break time-reversal symmetry (BTRS) on a square lattice as a function of applied stress.  
Calculations show that if critical temperatures coincide at zero stress, they exhibit a linear kink and no kink otherwise for uniaxial and isotropic strain. Linear kink is absent for shear strain.
We find that in general, the microscopic calculations show a complex phase diagram, for example, non-monotonic behavior of BTRS transition.
Another beyond-Ginzburg-Landau theory result is that $U(1)$ critical temperature can decrease under compressional [100] uniaxial strain for small Poisson ratio materials.
In the second part of the paper, we consider the effects of boundaries and finiteness of the sample on the strain-induced splitting of $T_c^{U(1)}$ and $T_c^{\mathbb{Z}_2}$ transitions.
A finite sample has BTRS boundary states with persistent superconducting currents over a wide range of band filling.
Overall, the BTRS dome occupies a larger band filling--temperature phase space region for a mesoscopic sample with [110] surface compared to an infinite system.
Hence, the presence of boundaries helps to stabilize the BTRS phase.
 
\end{abstract}

\maketitle

\section{Introduction}

Superconducting states with broken time-reversal symmetry [BTRS, $U(1) \cross \mathbb{Z}_2$] are sought-after states of matter, due to a plethora of phenomena, different from conventional superconductors that break only a single $U(1)$ gauge symmetry, and possible new applications.
In the paper, we focus on $s+id$ states.
Such states, for example were discussed  in connection with Sr$_2$RuO$_4$ \cite{grinenko2021split,ikegaya2021proposal, roising2022heat,romer2020theory} and iron-based superconductors \cite{grinenko2017superconductivity,grinenko2020superconductivity,grinenko2021state,lee2009pairing,platt2012mechanism,ghosh2025elastocaloric,liu2025evolution}.
An especially interesting question is how to control the time-reversal symmetry breaking.
It was proposed that $s+id$ states can appear in $d$-wave superconductors due to disorder, which can be experimentally controlled \cite{breio2022supercurrents,andersen2024spontaneous}.
Also disorder-induced $s+is$ states were discussed in  Refs.~\cite{bobkov2011time,stanev2014complex,silaev2017phase}.
The strain is a useful tool to control the time-reversal symmetry breaking.
Moreover, response to various strains 
is used to argue for or against breaking time-reversal symmetry in systems where this question is debated 
\cite{mattoni2025direct}.
In this paper, we investigate external stress as a method of controlling spontaneous breaking of time-reversal symmetry, in particular, $U(1) \cross \mathbb{Z}_2$ critical temperature.

\begin{figure}
    \begin{center}
		\includegraphics[width=0.8\columnwidth]{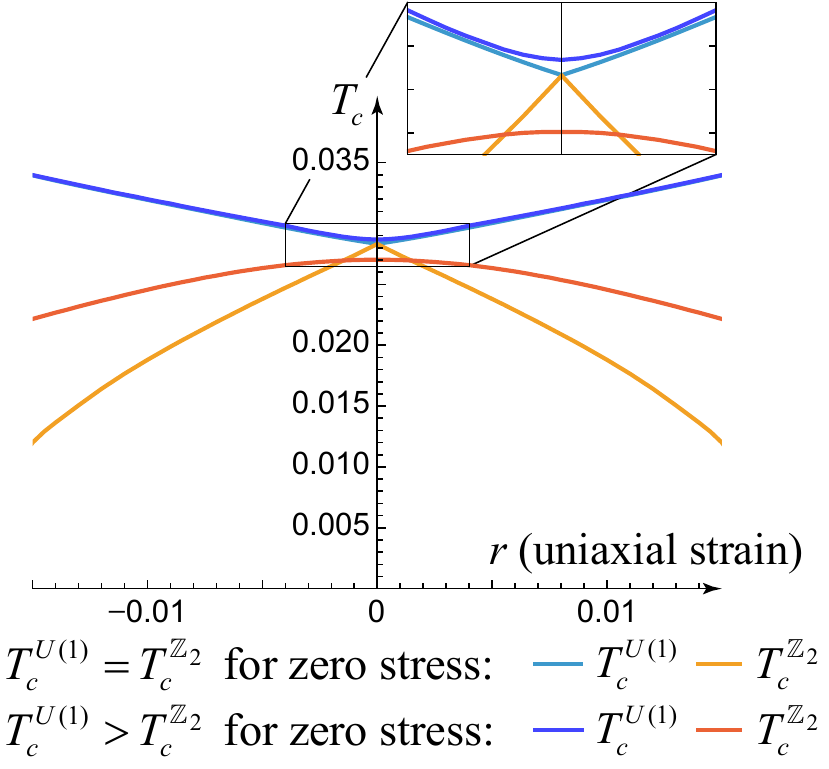}
		\caption{Superconducting and BTRS critical temperatures as a function of [100] uniaxial strain (a measure of orthorhombicity) within Ginzburg-Landau formalism. 
        Note the linear kink for $U(1)$ and $\mathbb{Z}_2$ critical temperatures at zero strain when system is tuned that $T_c^{U(1)}(0)=T_c^{\mathbb{Z}_2}(0)$.
        The kink is absent when $U(1)$ and $\mathbb{Z}_2$ critical temperatures are split at zero stress.
        Microscopic calculations are presented in \figref{fig:microscopic critical temperatures small stress}.}
		\label{fig:GL critical temperatures}
\end{center}
\end{figure}

External uniaxial stress leads to the change of superconducting critical temperature \cite{grinenko2021split,li2021high,barber2019role,li2022elastocaloric,barber2018resistivity,steppke2017strong,hicks2014strong,fischer2016fluctuation}.
More importantly, a combination of transport probes and muon spin relaxation ($\mu$SR) spectroscopy was interpreted as evidence for splitting of $U(1)$ and $\mathbb{Z}_2$ critical temperatures under external uniaxial stress in Sr$_2$RuO$_4$ \cite{grinenko2021split} and no splitting under isotropic stress \cite{grinenko2021unsplit}.
The behavior of $U(1)$ critical temperature in the vicinity of zero stress attracts a lot of attention for Sr$_2$RuO$_4$ \cite{mattoni2025direct,li2022elastocaloric,li2021high,hicks2014strong,steppke2017strong,jerzembeck2024t} and iron-based superconductors \cite{ghosh2025elastocaloric,liu2025evolution}. 
Ginzburg-Landau calculations of uniaxial stress predict the presence of a linear kink when critical temperatures coincide for zero strain (\figref{fig:GL critical temperatures}).
Otherwise, both critical temperatures behave continuously.

The [100] uniaxial strain $\varepsilon_{xx} - \varepsilon_{yy}$ fundamentally breaks the $x-y$ symmetry from the tetragonal $D_{4h}$ group down to the orthorhombic $D_{2h}$ group.
This leads to the fact that $d_{x^2-y^2}$-wave, extended $s$-wave ($s_{x^2+y^2}$), as well as isotropic $s$-wave, all belong to the same irreducible representation of the orthorhombic point group \cite{li1993mixed,shimahara2021stability,o1995mixed}. An intuitive explanation of this fact is illustrated in Ref.~\cite{li1993mixed}.
As a consequence, BTRS $s+id$ state has $\pi/2$ phase difference between components for the tetragonal system and $\in (0;\pi/2)$ phase difference for the orthorhombic one \cite{shimahara2021stability}.
A non-BTRS (just $U(1)$ symmetry breaking) ground state for the tetragonal system is either pure $s$-wave or pure $d$-wave, and for the orthorhombic system is $s+d$ or $s-d$ state.
In contrast, isotropic stress does not break the tetragonal $D_{4h}$ point group.
Shear $\varepsilon_{xy}$ and [110] uniaxial strain reduce the point group down to $D_{2h}$ (rotated by $\pi/4$).
In these three cases, $d_{x^2-y^2}$-wave and extended $s$-wave belong to different irreducible representations.
Hence, BTRS $s+id$ state forms with $\pi/2$ phase difference.

We study a system that has coinciding $U(1)$ and $\mathbb{Z}_2$ critical temperatures for unstrained case, and $T_c^{U(1)}$ and $T_c^{\mathbb{Z}_2}$ split under external stress.
We investigate the general properties of the splitting for a square lattice with nearest-neighbor hopping.
The theoretical description can be done using Ginzburg-Landau (GL) theory \cite{li1993mixed,volovik1988splitting,hess1989broken,yuan2021strain,fischer2016fluctuation} or microscopic theories \cite{shimahara2021stability,jurecka1999phase,musaelian1996mixed,o1995mixed,o1995s,ghosh1999two,romer2020theory,yuan2021strain}.
A standard GL approach predicts an increase of $U(1)$ superconducting critical temperature and a decrease of $T_c^{\mathbb{Z}_2}$ under external [100] uniaxial stress (see \figref{fig:GL critical temperatures} and \appref{app:GL method} for the computational method) \cite{li1993mixed}.
The first fact is easy to show (see \appref{app:GL method}).
However, our microscopic calculation, including the Poisson effect, shows that very different scenarios are possible.
Namely, we consider the appearance of transverse strain in response to the longitudinal strain and show that $T_c^{U(1)}$ can increase or decrease depending on the Poisson ratio.
Moreover, we show that in fact $T_c^{\mathbb{Z}_2}$ can have nonmonotonic dependence on strain in some Poisson ratio region.
The microscopic approach allows us to calculate phase diagrams outside the validity of the standard GL approach.
This is important since the nonmonotonic dependence of $T_c^{\mathbb{Z}_2}$ on strain happens when the difference between $U(1)$ and $\mathbb{Z}_2$ critical temperatures can not be assumed to be small.

The peculiar behavior of $\mathbb{Z}_2$ critical temperature on [100] uniaxial strain is observed in particular for Poisson ratio $\nu$ around 0.3.
The decrease (or approximate independence) of $U(1)$ critical temperature on strain happens for $\nu \lesssim 0.25$.
Such parameters are similar to those of, for example, iron-based superconductors that have a Poisson ratio in the range [0.18; 0.36] \cite{imai2017comparative}.
However, some iron-based compounds have proximity to van Hove singularities under strain \cite{phan2017effects,barber2019role}.
This can overcome the effect discussed in the paper.

The paper is structured as follows: First, we introduce general formalism in \secref{sec:general model}.
Then, we present numerical superconducting phase diagrams and results on $T_c^{U(1)}$ and $T_c^{\mathbb{Z}_2}$ splitting under [100] uniaxial stress for nearest-neighbor pairing interaction in \secref{sec:nearest neighbor}.
Section \ref{sec:additional onsite interaction} contains similar plots for a model with additional on-site interaction potential, exploring the robustness of the results with respect to changes in the model.
Section \ref{sec:shear and isotropic} contains results on critical temperatures split under shear and isotropic strain.
Phase diagram calculation results for a strained finite system with [110] surface, which can host boundary BTRS states, are illustrated in \secref{sec:finite system}.
Conclusion and perspectives are presented in \secref{sec:conclusion}.

\section{The model for uniaxial strain} \label{sec:general model}

In a standard approximation (considering a single electronic band), a superconducting gap can be written as a sum of different irreducible representations \cite{o1995mixed}
\begin{equation} \label{eq:gap general}
    \Delta(\bk) = \sum_{\alpha}{\Delta_\alpha \gamma_\alpha(\bk)},
\end{equation}
where $\gamma_\alpha(\bk)$ are principal basis functions of symmetries $\alpha$.
The basis functions are not necessarily orthogonal (for instance, when considering on-site and nearest-neighbor interactions).
The interaction potential $V(\bk, \bk')$ in Fourier space reads as
\begin{equation} \label{eq:interaction general}
    V(\bk, \bk') = \sum_{\alpha}{V_\alpha \gamma_\alpha (\bk) \gamma_\alpha (\bk')}.
\end{equation}
The self-consistency gap equation is
\begin{equation} \label{eq:self-consistency for total gap general}
    \Delta(\bk) =  \int_{\bz} \frac{d\bk'}{S_{\bz}} V(\bk, \bk') \Delta(\bk') \frac{\tanh{\frac{E_{\bk'}}{2T}}}{2E_{\bk'}},
\end{equation}
where $S_{\bz}$ is the area of the first Brillouin zone (BZ), $E_{\bk} = \sqrt{\xi (\bk)^2 + |\Delta (\bk)|^2}$ is quasiparticle excitation energy, $\xi (\bk)$ is an electron dispersion relation, and $k_B = 1$.
Combining Eqs.~(\ref{eq:gap general})--(\ref{eq:self-consistency for total gap general}), one obtains self-consistency equation for superconducting gap representations
\begin{equation} \label{eq:self-consistency components}
    \Delta_\alpha = \sum_{\beta}{S_{\alpha \beta} \Delta_\beta}
\end{equation}
with
\begin{equation} \label{eq:matrix S}
    S_{\alpha \beta} = V_\alpha \int_{\bz} \frac{d\bk'}{S_{\bz}} \gamma_\alpha (\bk') \gamma_\beta (\bk') \frac{\tanh{\frac{E_{\bk'}}{2T}}}{2E_{\bk'}}.
\end{equation}
It is a set of coupled integral equations.
It has a nontrivial solution ($\Delta(\bk) \neq 0$) when the matrix $S$ has the largest eigenvalue equal to unity and a trivial solution ($\Delta(\bk) = 0$) otherwise.
If the matrix $S$ has two the largest eigenvalues equal to unity, it corresponds to the BTRS state in this model.
Note that matrix $S$ is symmetric only in the case when all $V_\alpha$ are the same.
It is not always the case, for instance, we investigate the situation in \secref{sec:additional onsite interaction}. 
An alternative viewpoint on the phase transitions can be obtained from the analysis of the Jacobian of \eqref{eq:self-consistency components} \cite{spathis1992perturbed}.
Each phase transition results from pitchfork bifurcations (when the Jacobian becomes singular).
In the paper, we use an eigenvalue approach and numerical methods to find eigenvalues of $S$.

Orthorhombic deformation of the crystal lattice due to [100] uniaxial stress affects both dispersion relation $\xi(\bk)$ and pairing interaction $V(\bk, \bk')$.
However, we retain changes only in $\xi(\bk)$ due to the strain to capture the physics of the phenomenon qualitatively:
\begin{equation} \label{eq:dispersion relation}
    \xi(\bk) = -2 (1+ \delta t) \cos k_x - 2 (1 - \nu \delta t) \cos k_y - \mu.
\end{equation}
Here $\delta t$ ($- \nu \delta t$) is the change of hopping integral in the $x$ ($y$) direction due to [100] uniaxial stress, $\nu$ is the Poisson ratio.
We investigate $\nu \in [0;1]$.
It corresponds to the range from no $y$ strain under external stress in $x$ direction ($\nu =0$) to a perfect system with $xy$-plane area conservation under external stress ($\nu=1$).
The linear dependence of hopping integrals on uniaxial external stress was confirmed experimentally for Sr$_2$RuO$_4$ \cite{sunko2019direct}.
Their setup allowed us to check it for the anisotropic strain $|\varepsilon_{xx} - \varepsilon_{yy}|$ up to $1.7 \%$, which corresponds to a relative change of hopping by $10 \%$.
Therefore, throughout the paper, we use the terms stress, strain, and $\delta t$ as synonyms.
An approximation similar to \eqref{eq:dispersion relation} was used to discuss various aspects of phase diagrams in Refs.~\cite{shimahara2021stability,jurecka1999phase,o1995mixed,o1995s,yuan2021strain} (with $\nu=0$) in contrast to \cite{ghosh1999two,liu1997mixed} who varied both dispersion and interaction.

\section{The case of nearest-neighbor interaction under uniaxial strain} \label{sec:nearest neighbor}

In this section, we consider spin-singlet pairing potential arising purely from nearest-neighbor interactions:
\begin{equation} \label{eq:interaction 2 components}
    V(\bk, \bk') = V_1 \left[ \gamma_d (\bk) \gamma_d (\bk') + \gamma_{\sext} (\bk) \gamma_{\sext} (\bk') \right].
\end{equation}
It corresponds to the two gap representations
\begin{equation} \label{eq:gap and basis functions 2 components}
\begin{gathered}
    \Delta(\bk) = \Delta_{d} \gamma_d (\bk) + \Delta_{\sext} \gamma_{\sext} (\bk), \\
    \gamma_d (\bk) = \cos k_x - \cos k_y, \quad \gamma_{\sext} (\bk) = \cos k_x + \cos k_y,
\end{gathered}
\end{equation}
where the basis function $\gamma_d (\bk)$ corresponds to $d_{x^2-y^2}$-wave (later $d$-wave, $B_{1g}$ irreducible representation), and $\gamma_{\sext}$ corresponds to extended $s_{x^2+y^2}$-wave (later $\sext$-wave, $A_{1g}$ irreducible representation).

The matrix $S$ [\eqref{eq:matrix S}] is a symmetric $2 \times 2$ matrix for the case.
BTRS state requires double eigenvalue degeneracy ($\lambda_1 = \lambda_2 = 1$).
A necessary and sufficient condition for this is to have unit diagonal terms and zero off-diagonal terms.
Zero off-diagonal terms have been shown to correspond to an energetically stable BTRS solution \cite{shimahara2021stability}.
It is guaranteed for an undistorted system ($\delta t = 0$) by symmetry of $E_{\bk}$ and antisymmetry of $\gamma_d (\bk) \gamma_{\sext} (\bk)$ w.r.t. $k_x \leftrightarrow k_y$.
This results in a $\pi/2$ phase difference between superconducting gap components.
There are three different superconducting phases in the case ($\delta t =0$): Pure $\sext$-wave, pure $d$-wave, and $\sext + id$ phase.
Transitions between the phases are of the second order at least at the mean-field level in these models \cite{musaelian1996mixed,shimahara2021stability}.
Off-diagonal elements are not necessarily zero for an orthorhombic system.
This leads to a mixed $\sext \pm d$ state with 0 or $\pi$ phase difference between gap representations.
The BTRS state can remain stable when the relative phase shifts so that off-diagonal elements nullify.
This fact was recently discussed in detail in \cite{shimahara2021stability}, and phase difference $\in (0; \pi/2)$ was previously observed in numerical calculations for orthorhombic systems \cite{jurecka1999phase,o1995mixed,o1995s}.

\begin{figure}[h]
    \begin{center}
		\includegraphics[width=0.99\columnwidth]{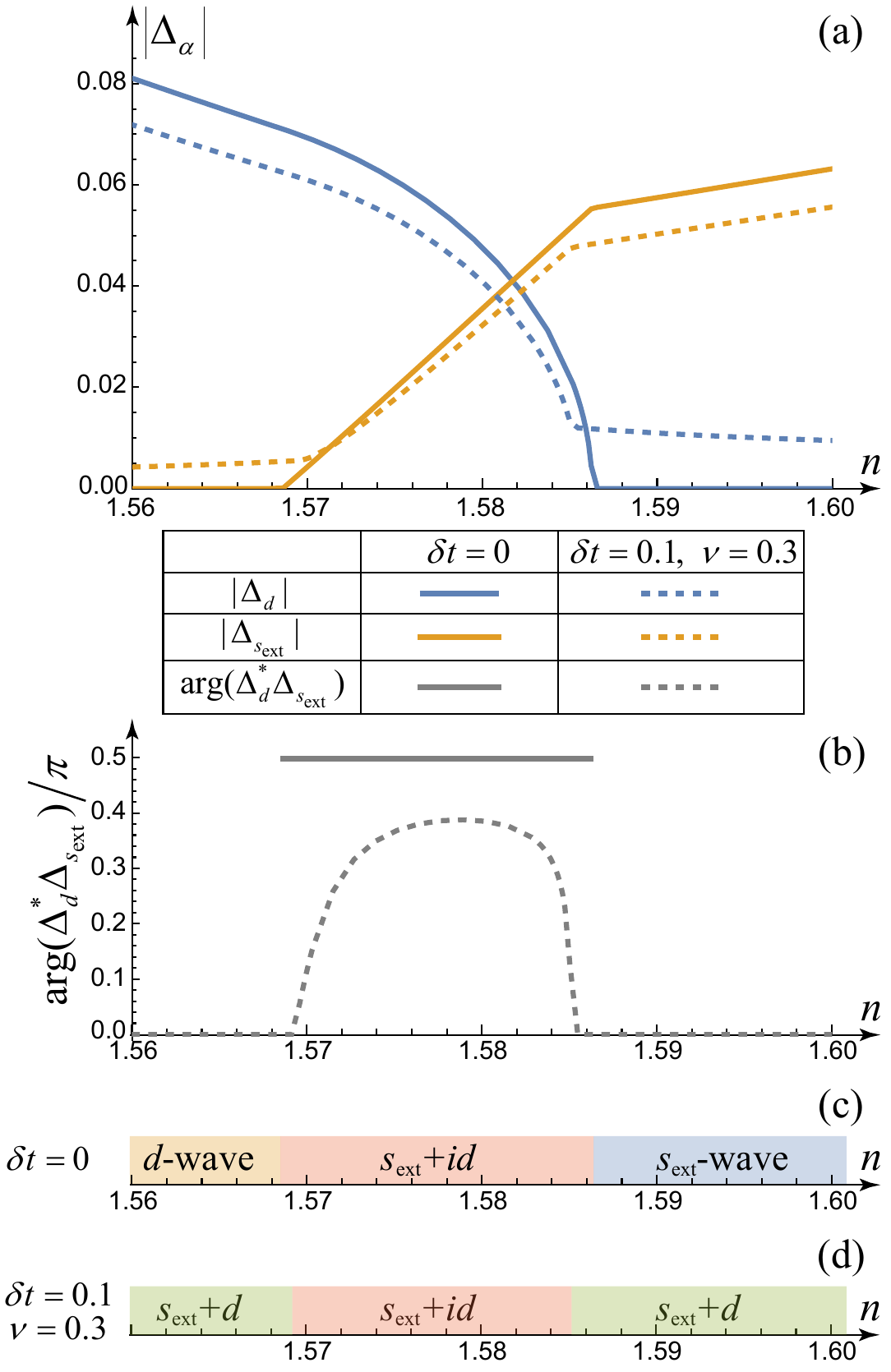}
		\caption{(a) The dependence of the superconducting gap components $|\Delta_\alpha|$ on the band filling $n$ for the model with gap $ \Delta(\bk) = \Delta_{d} (\cos k_x - \cos k_y) + \Delta_{\sext} (\cos k_x + \cos k_y)$.
        The blue (orange) line corresponds to the $d$-wave ($\sext$-wave) gap irreducible representation.
        (b) The dependence of the phase difference between gap components $\arg (\Delta_d^* \Delta_{\sext})$ on the band filling $n$.
        Solid and dashed lines correspond to the tetragonal (change of hopping integral $\delta t = 0$) and orthorhombic ([100] uniaxial strain $\delta t = 0.1$, Poisson ratio $\nu=0.3$, $\delta t_y = -\nu \delta t= -0.03$) system, respectively.
        (c) Phase diagram for an unstrained system.
        The $s+id$ state has $\pi/2$ phase difference between components.
        (d) Phase diagram for orthorhombic ($\delta t = 0.1$, $\nu=0.3$) system.
        Both gap components are always non-zero.
        The $\sext+d$ state has coinciding phases of both components.
        The $\sext+id$ state has phase difference $\in(0;\pi/2)$ between components.
        Nearest-neighbor interaction strength $V_1=2$, $T=0$.}
		\label{fig:gaps and phase difference 2 components}
\end{center}
\end{figure}

\subsection*{Numerical results}

We start by illustrating the components of the superconducting gap and the relative phase between them as a function of the band filling $n$ at zero temperature (\figref{fig:gaps and phase difference 2 components}).
Self-consistency \eqref{eq:self-consistency components} is satisfied for an extremum of Gibbs free energy.
We fix chemical potential $\mu$ and temperature $T$, solve \eqref{eq:self-consistency components} for gaps, and then compute corresponding electron density $n$ (repeating the whole procedure for different $\mu$).
We plot results for the tetragonal undistorted system ($\delta t = 0$) and the system under [100] uniaxial stress ($\delta t = 0.1$, $\nu=0.3$).
One observes BTRS, pure $\sext$, and $d$-wave phases for undistorted systems.
The phase difference is $\pi/2$ in the BTRS phase and not indicated outside the BTRS region (because one of $\Delta_\alpha = 0$) for $\delta t = 0$.
Gap changes continuously in $\sext + id$ region that is consistent with a second-order phase transition at the mean-field level.
On the contrary, when tetragonal symmetry is broken, the ground state is a BTRS phase or $\sext + d$ phase.
Both gap components $\Delta_d, \, \Delta_{\sext}$ are always nonzero.
The phase difference changes continuously in the BTRS phase for the orthorhombic system and is less than $\pi/2$ (relevant for an undistorted system).
The picture is typical for all Poisson ratios and relatively small distortions ($\delta t < 0.2$).
We numerically checked that both eigenvalues of $S$ become unity in the BTRS phase and off-diagonal terms become zero for the orthorhombic system.

\begin{figure}[!h]
    \begin{center}
		\includegraphics[width=0.84\columnwidth]{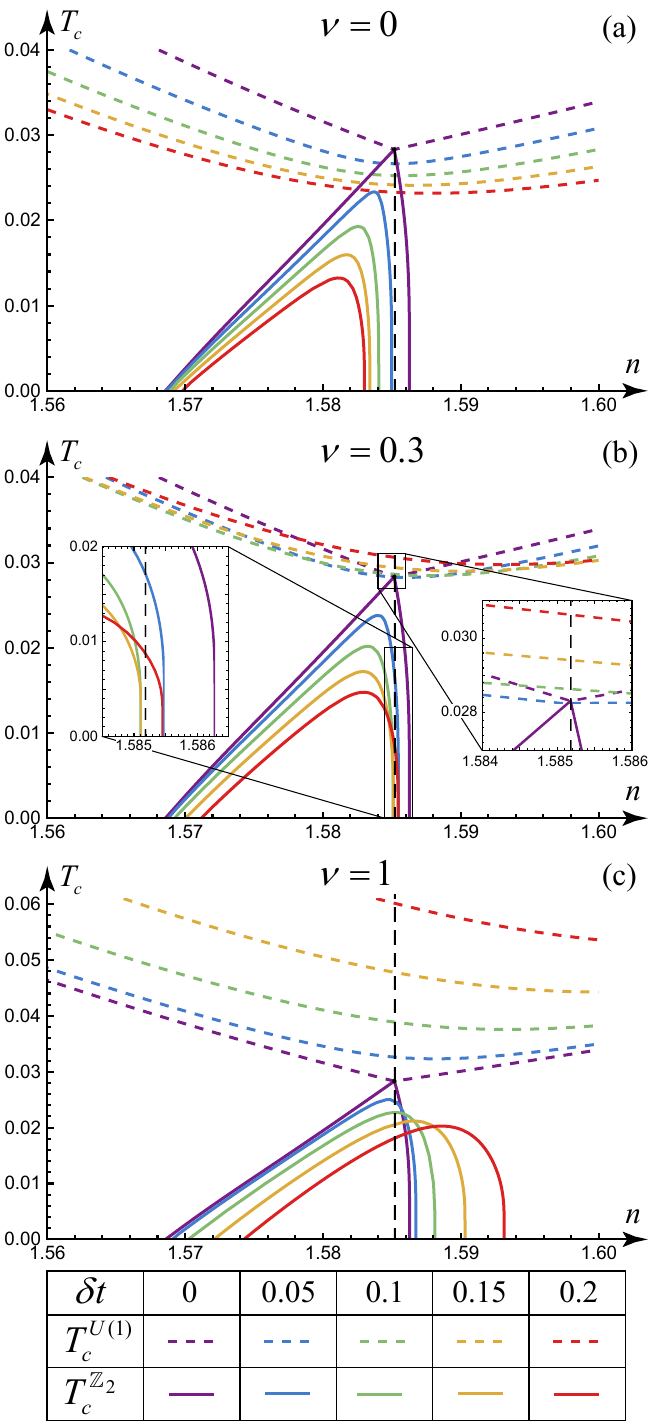}
		\caption{Superconducting phase diagrams in band filling $n$, temperature $T$ coordinates for different measures of orthorhombicity $\delta t$ for Poisson ratios
        (a) $\nu=0$, (b) $\nu=0.3$, (c) $\nu=1$.
        Dashed and solid lines correspond to the $U(1)$ and $\mathbb{Z}_2$ symmetry-breaking critical temperatures, respectively.
        Above the dashed line is a normal metal phase, between the dashed and solid line is $U(1)$ symmetry-breaking phase, and below the solid line is the BTRS phase.
        BTRS dome moves monotonously to a smaller (larger) band filling under stress for $\nu=0$ ($\nu=1$).
        BTRS dome first moves towards lower and then towards a larger band filling under stress for $\nu=0.3$.
        Vertical black dashed line corresponds to band filling $n=1.58518$ where $T_c^{U(1)}=T_c^{\mathbb{Z}_2}$ for tetragonal system ($\delta t = 0$).
        Nearest-neighbor interaction strength $V_1=2$.}
		\label{fig:phase diagrams 2 components}
\end{center}
\end{figure}

Next, we analyze the phases with additional variation of temperature.
Phase diagrams in coordinates $(n,T)$ for different Poisson ratios ($\nu =0, \, 0.3, \, 1$) are presented in \figref{fig:phase diagrams 2 components}.
The criterion for $T_c^{U(1)}$ is the maximal eigenvalue of $S$ equals to unity and gap components $\Delta_d, \, \Delta_{\sext} \rightarrow 0$.
The criterion for $T_c^{\mathbb{Z}_2}$ is that both eigenvalues of $S$ are unity and the phase difference between gap components is 0 or $\pi$.
Above the dashed lines ($T_c^{U(1)}$) is a normal metal phase.
Below the solid lines is the BTRS superconducting phase.
There is a pure $d$-wave ($\sext$-wave) phase below the dashed purple line and to the left (right) of the purple BTRS dome for a tetragonal system.
Note the steep right side of the BTRS dome ($\sext+id$ to $d$-wave transition).
The origin of the property for the tetragonal system is explained in Ref.~\cite{talkachov2025type15}.
Orthorhombic system ($\delta t \neq 0$) has $\sext+d$ phase below the dashed line and beyond the BTRS (solid line) dome.
It confirms that the properties mentioned in the previous paragraph can be extended to finite temperatures.
Increasing the orthorhombicity ($\delta t$) leads to decreasing the height of the BTRS dome [$\max (T_c^{\mathbb{Z}_2})$] for all $\nu$.
Increasing the orthorhombicity ($\delta t$) also leads to shrinking the BTRS dome horizontal size (band filling range) for $\nu =0, \, 0.3$.
However, the horizontal size of the BTRS dome increases for $\nu = 1$ when $\delta t$ rises.
Another peculiar property is that the BTRS dome moves towards a smaller (larger) band filling $n$ under external stress for $\nu =0$ ($\nu = 1$).
In contrast, first, the BTRS dome moves to the smaller band filling and then (for $\delta t \gtrsim 0.12$) moves to larger $n$ for $\nu = 0.3$ when $\delta t$ increases.
This nonmonotonic behavior is observed for systems with the Poisson ratio approximately in the range $[0.25;0.4]$.

\begin{figure}
    \begin{center}
		\includegraphics[width=0.99\columnwidth]{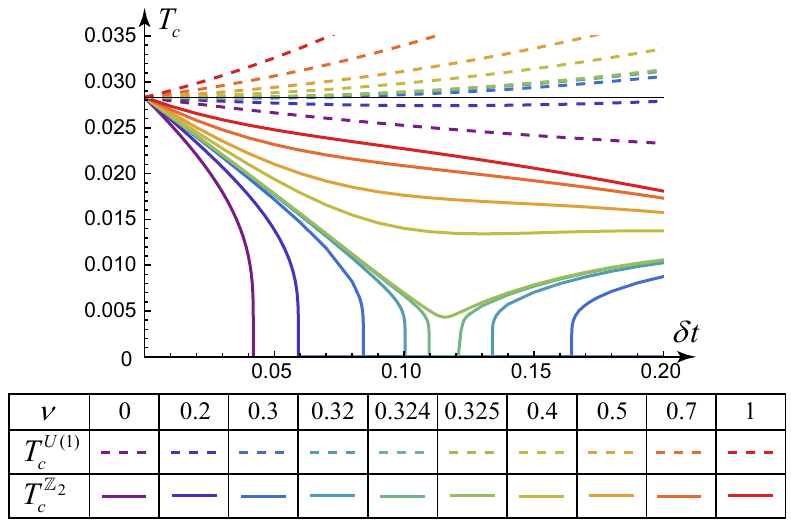}
		\caption{Superconducting and BTRS critical temperatures as a function of change of hopping $\delta t$ (a measure of orthorhombicity) due to external [100] uniaxial compressional stress for different Poisson ratios $\nu$.
        Dashed and solid lines correspond to the $U(1)$ and $\mathbb{Z}_2$ symmetry-breaking critical temperatures, respectively.
        Ginzburg-Landau behavior of critical temperatures in \figref{fig:GL critical temperatures} qualitatively describes microscopic calculations only for $\nu=1$.
        The horizontal black line indicates the critical temperature for the tetragonal system without stress $T_c^{U(1)}=T_c^{\mathbb{Z}_2}=0.02832$, $\delta t =0$, band filling $n=1.58518$.
        Nearest-neighbor interaction strength $V_1=2$.}
		\label{fig:Tc stress 2 components}
\end{center}
\end{figure}

\begin{figure}
    \begin{center}
		\includegraphics[width=0.8\columnwidth]{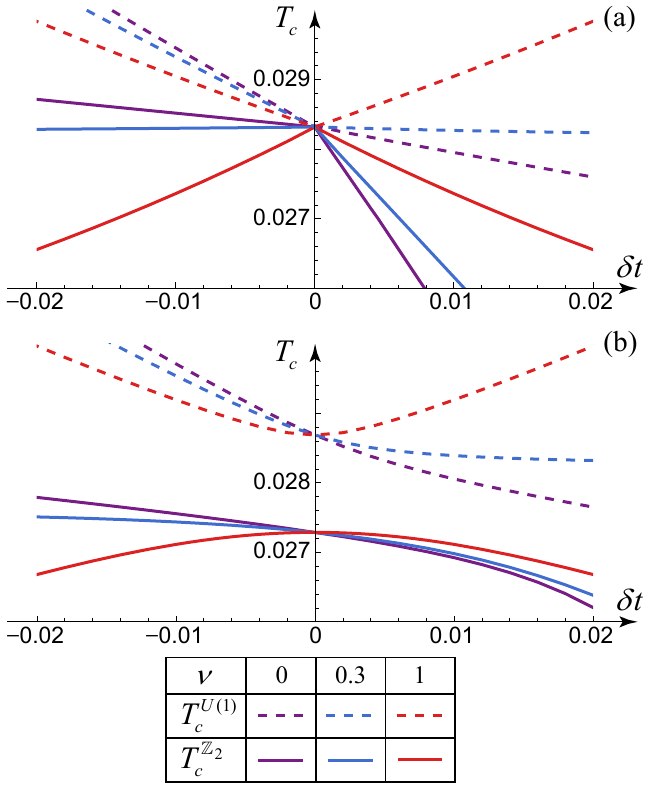}
		\caption{Superconducting and BTRS critical temperatures as a function of tensile ($\delta t <0$) and compressional ($\delta t >0$) [100] uniaxial strain within microscopic model. 
        (a) System is tuned that $T_c^{U(1)}=T_c^{\mathbb{Z}_2}$ at zero external strain, band filling $n=1.58518$.
        Note the linear kink for $U(1)$ and $\mathbb{Z}_2$ critical temperatures at zero strain.
        (b) System has $T_c^{U(1)}>T_c^{\mathbb{Z}_2}$ at zero external strain, $n=1.58458$.
        Note the absence of kink in critical temperatures at zero stress for the case (b).
        Compressive and tensile stress are not symmetric for $\nu \neq 1$.}
		\label{fig:microscopic critical temperatures small stress}
\end{center}
\end{figure}

Let us focus on a case with the band filling $n=1.58518$ (black dashed line in \figref{fig:phase diagrams 2 components}).
It corresponds to an undistorted system that has coinciding critical temperatures ($T_c^{U(1)}=T_c^{\mathbb{Z}_2}=0.02832$).
It means a direct transition from normal metal to $U(1) \cross \mathbb{Z}_2$ superconductor.
This is a tetracritical point on the $(n,T)$ diagram that attracted significant interest due to the physics of topological defects  \cite{yuan2021strain,talkachov2025type15}.
Also, such points are interesting from the viewpoint of the physics of fluctuations.
Namely,  going beyond mean-field, this position on the phase diagram is especially favorable for fluctuations
to switch the sequence of the phase transition, leading
to time-reversal symmetry breaking above the superconducting state \cite{grinenko2021state,bojesen2014phase}.
The situation ($T_c^{U(1)}\approx T_c^{\mathbb{Z}_2}$) is reported in Sr$_2$RuO$_4$ as a consequence of $\mu$SR measurements \cite{grinenko2021split,xia2006high}.
One can find $U(1)$ and $\mathbb{Z}_2$ critical temperatures for different stress values ($\delta t$) from the intersection of colored lines with the black dashed line in \figref{fig:phase diagrams 2 components}.
Band filling $n$ remains constant under crystal deformation.
This results in a critical temperature as a function of orthorhombicity parameter $\delta t$ plot illustrated in Figs.~\ref{fig:Tc stress 2 components}, \ref{fig:microscopic critical temperatures small stress}(a).
We plot critical temperatures only for a few Poisson ratio values.
Though we computed $T_c$'s for many other $\nu$ values, and the analysis below is based on the extended dataset.
The $U(1)$ critical temperature (dashed lines) has a monotonic decreasing (increasing) trend for compressive strain ($\delta t > 0$) for $\nu \in [0;0.1]$ ($\nu \in [0.35;1]$).
However, systems with an intermediate Poisson ratio approximately in the range $(0.1;0.35)$ first have a decrease with the following increase in $T_c^{U(1)}$.
This contradicts the GL theory result that $T_c^{U(1)}$ should always increase when orthorhombicity increases \cite{li1993mixed} (see also \appref{app:GL method} in this paper).

Next, we focus on the $\mathbb{Z}_2$ critical temperature.
The BTRS critical temperature (solid lines in \figref{fig:Tc stress 2 components}) is a slowly monotonously decreasing function of strain $\delta t$ for Poisson ratio $\nu \in (0.45;1]$.
However, $\mathbb{Z}_2$ critical temperature shows much more complex behavior: We find that it is a non-monotonous function for Poisson ratio $\nu \in (0.25;0.45)$.
Moreover, BTRS critical temperature decreases to zero and then reappears and grows for larger external stress $\delta t$ for materials with $\nu \in (0.25;0.3246)$.
This behavior is possible to understand on the example of $\nu=0.3$ using \figref{fig:phase diagrams 2 components}(b) left inset by tracking BTRS dome displacement due to stress $\delta t$.
Also note the sharp drop and abrupt reentrance of $T_c^{\mathbb{Z}_2}$ in \figref{fig:Tc stress 2 components}.
This has its origin from the steep right side of the BTRS dome (\figref{fig:phase diagrams 2 components}).
Namely, a small shift to the left of the BTRS dome leads to significant vertical displacement (change of $T_c$) of its intersection with the vertical $n=1.58518$ line.

We saw from microscopic calculations that the phase diagram shows much more complex strain-temperature dependence than one that follows from the simplest GL theory.
In particular, a significant difference occurs when the Poisson ratio $\nu$ is different from unity.
Microscopic results for $\nu=1$ are qualitatively similar to critical temperature behavior in GL formalism (\figref{fig:GL critical temperatures}).
The difference in thermal contraction for substrate and sample with fine-tuned $T_c^{U(1)}\approx T_c^{\mathbb{Z}_2}$ leads to thermally-induced pre-strains that split $U(1)$ and $\mathbb{Z}_2$ critical temperatures \cite{mattoni2025direct}.
Also situation with $T_c^{U(1)}>T_c^{\mathbb{Z}_2}$ can be achieved by doping \cite{ghosh2025elastocaloric}.
Let us consider the case of band filling $n=1.58458$ corresponding to the situation.
In this case, $U(1)$ symmetry-breaking phase corresponds to pure $d$-wave at zero stress.
Strain dependence of critical temperatures is illustrated in \figref{fig:microscopic critical temperatures small stress}(b).
Note the asymmetry of critical temperatures w.r.t. change of compressional to tensile stress (for Poisson ratio $\nu \neq 1$, see \figref{fig:microscopic critical temperatures small stress} and plots in \appref{app:additional plots}).
In particular, zero strain does not correspond to a minimum of $U(1)$ critical temperature for microscopic calculations with $\nu \neq 1$.
The asymmetry is a common feature in experimental results on $U(1)$ critical temperature strain dependence \cite{ghosh2025elastocaloric,hicks2014strong,steppke2017strong,liu2025evolution,mattoni2025direct}.

This section naturally leads to the question: Is it this particular microscopic model that produces such peculiar results, or is it a more general property for this family of systems?
We add on-site interaction potential $V_0$ to the model of this section to address the question, and in the next section, focus mostly on the main point of interest: The $\mathbb{Z}_2$ critical temperature.

\section{Additional on-site interaction} \label{sec:additional onsite interaction}

In this section, we consider spin-singlet pairing potential arising from on-site $V_0$ and nearest-neighbor $V_1$ interactions:
\begin{equation} \label{eq:interaction 3 components}
    V(\bk, \bk') = V_0 + V_1 \left[ \gamma_d (\bk) \gamma_d (\bk') + \gamma_{\sext} (\bk) \gamma_{\sext} (\bk') \right].
\end{equation}
It corresponds to the following three gap components,
\begin{equation} \label{eq:gap and basis functions 3 components}
    \Delta(\bk) = \Delta_{s} + \Delta_{d} \gamma_d (\bk) + \Delta_{\sext} \gamma_{\sext} (\bk),
\end{equation}
where we implicitly used $\gamma_s (\bk) = 1$ for the uniform gap in momentum space.

The $3 \times 3$ matrix $S$ [\eqref{eq:matrix S}] is not symmetric for a distorted system (except the case $V_0 = V_1$).
It becomes a block diagonal matrix (the blocks are $1 \times 1$ for $d$-wave and $2 \times 2$ for two components of $s$-wave) in the tetragonal system ($\delta t = 0$).
Therefore, superconducting phases are the following: Pure $d$-wave, non-BTRS mixed $s$-wave ($s+\sext$ or $s-\sext$ state), and BTRS state.
The BTRS state has $\pm \pi / 2$ phase difference between $d$-wave and $s$-waves and  0 or $\pi$ phase difference between $s$-waves.
The picture is more complicated for the orthorhombic system ($\delta t \neq 0$) and will be addressed below.
In general, we expect to have all three gap components $\Delta_{s}, \, \Delta_{d}, \, \Delta_{\sext}$ with non-vanishing amplitude.
It follows from the fact that all matrix elements $S_{\alpha \beta}$ are nonzero.
A $U(1)$ symmetry-breaking state would correspond to 0 or $\pi$ phase difference between these components, and the opposite for BTRS state.

\begin{figure}
    \begin{center}
		\includegraphics[width=0.99\columnwidth]{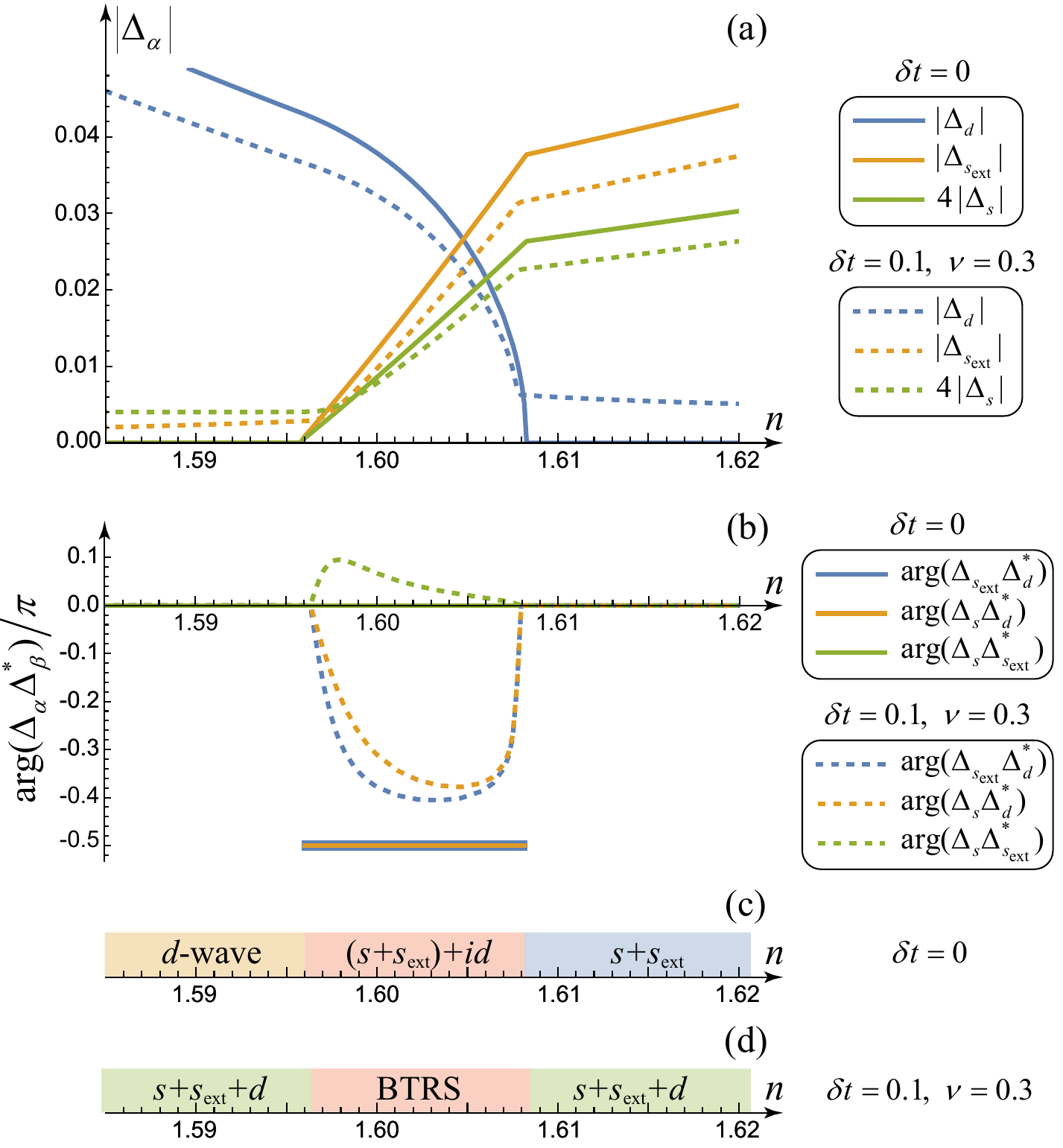}
		\caption{(a) The dependence of the superconducting gap components $|\Delta_\alpha|$ on the band filling $n$ for the model with gap $ \Delta(\bk) = \Delta_{s} + \Delta_{d} (\cos k_x - \cos k_y) + \Delta_{\sext} (\cos k_x + \cos k_y)$.
        Blue, orange, and green lines correspond to $d$-wave, $\sext$-wave, and $s$-wave gap irreducible representations, respectively.
        (b) The dependence of the phase difference between gap components $\arg (\Delta_\alpha \Delta_\beta^*)$ on the band filling $n$.
        Solid and dashed lines correspond to the tetragonal (change of hopping integral $\delta t = 0$) and orthorhombic ($\delta t = 0.1$, Poisson ratio $\nu=0.3$, $\delta t_y = -\nu \delta t= -0.03$) system, respectively.
        (c) Phase diagram for an unstrained system.
        BTRS state has coinciding phases of $s$-wave and $\sext$-wave gap components; $d$-wave has a phase that differs by $\pi/2$.
        (d) Phase diagram for orthorhombic ($\delta t = 0.1$, $\nu=0.3$) system.
        All gap components are always non-zero.
        They have coinciding phases for $s+\sext+d$ state and a non-zero phase difference between all components in BTRS state.
        On-site and nearest-neighbor interaction strengths are $V_0 = -0.5$ and $V_1=2$, respectively, $T=0$.}
		\label{fig:gaps and phase difference 3 components}
\end{center}
\end{figure}

\subsection*{Numerical results}

Again, we start from the characterization of superconducting gap components and phases between them at zero temperature (\figref{fig:gaps and phase difference 3 components}).
The tetragonal system has three superconducting phases as discussed in the previous paragraph: $d$-wave, $(s+\sext)$-wave (zero phase difference between components), and $(s+\sext)+id$ phase.
The gap for $s$-wave symmetry is significantly smaller (approximately 6 times) than for $\sext$-wave.
This can be partially explained by the significant difference in the interaction strength we chose: $V_0=-0.5$ and $V_1=2$.
Applying external stress leads to the two superconducting phases: $s+\sext+d$ state (zero phase difference between all components) and BTRS state, where all the phase differences are not multiple of $\pi$.
This picture is typical for all values of the on-site interaction $V_0 < 0$ we calculated.
We observe $\sext-s+d$ state around the BTRS dome for $V_0 > 0$ (plots are not presented in the paper).
The transition to the BTRS state is of the second order at the mean-field level.

The appearance of the nontrivial phase difference between $s$-wave and $\sext$-wave in the BTRS phase for the orthorhombic system can be seen from \eqref{eq:self-consistency components}.
For the three gap components [\eqref{eq:gap and basis functions 3 components}] assuming $U(1)$ gauge that fixes $\arg \Delta_d=0$ (measuring phases of two $s$-wave components relative to the $d$-wave) one gets
\begin{equation} \label{eq:nontrivial phases 1}
    \begin{split}
    \cot (\arg \Delta_{\sext}) - \cot (\arg \Delta_s) &= \frac{S_{\sext d} \Delta_d}{S_{\sext s} \text{Im} \Delta_s} \\
    &= \frac{- S_{s d} \Delta_d}{S_{s \, \sext} \text{Im} \Delta_{\sext}}.
\end{split}
\end{equation}
The phase difference ($\arg \Delta_{\sext} - \arg \Delta_s$) is non-zero for $S_{s d}\neq 0$ in the BTRS phase (when $\text{Im} \Delta_{\sext} \neq 0$).
Here $S_{s d}$ describes the coupling of $s$-wave to $d$-wave that becomes non-zero under orthorhombic distortion.

Now let us explain the behavior of phase difference $\arg \left(\Delta_s \Delta_{\sext}^*\right)$ as a function of band filling [see green dashed line in \figref{fig:gaps and phase difference 3 components}(b)].
Equation (\ref{eq:nontrivial phases 1}) can be rewritten as
\begin{equation}
    \arg \left(\Delta_s \Delta_{\sext}^*\right) \approx \frac{S_{\sext d}}{S_{\sext s}} \frac{\Delta_d}{|\Delta_s|} \sin (\arg \Delta_s).
\end{equation}
Here we used small phase difference $\arg \left(\Delta_s \Delta_{\sext}^*\right)$ approximation.
Matrix elements $S_{\alpha \beta}$ stay approximately constant in the whole BTRS phase (for fixed $T$).
Ratio $\Delta_d / |\Delta_s|$ is a monotonically decreasing function of band filling.
It drops by a factor of fifty from LHS to RHS of BTRS dome in \figref{fig:gaps and phase difference 3 components}(a).
Function $\sin (\arg \Delta_s)$ is bell-shaped in BTRS region [see $\arg (\Delta_s \Delta_d^*)$ behavior in \figref{fig:gaps and phase difference 3 components}(b)].
Therefore, phase difference $\arg \left(\Delta_s \Delta_{\sext}^*\right)$ should be a bell-shaped function with maximum shifted towards large values of $\Delta_d / |\Delta_s|$ (to the small band filling).

\begin{figure}[t]
    \begin{center}
		\includegraphics[width=0.99\columnwidth]{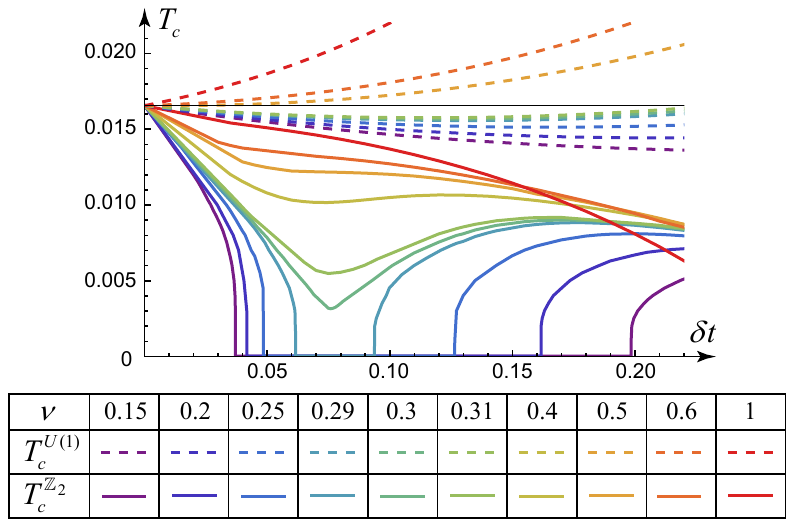}
		\caption{Superconducting and BTRS critical temperatures as a function of change of hopping $\delta t$ (a measure of orthorhombicity) due to external [100] uniaxial compressional stress for different Poisson ratios $\nu$.
        Dashed and solid lines correspond to the $U(1)$ and $\mathbb{Z}_2$ symmetry-breaking critical temperatures, respectively.
        Ginzburg-Landau behavior of critical temperatures in \figref{fig:GL critical temperatures} qualitatively describes microscopic calculations only for $\nu=1$.
        General behavior is similar to \figref{fig:Tc stress 2 components} with only nearest-neighbor pairing.
        Hence, the complex beyond Ginzburg-Landau description behavior persists over a certain generalization of the model.
        The horizontal black line indicates the critical temperature for the tetragonal system without stress $T_c^{U(1)}=T_c^{\mathbb{Z}_2}=0.01656$, $\delta t =0$, band filling $n=1.60777$.
        On-site and nearest-neighbor interaction strengths are $V_0 = -0.5$ and $V_1=2$, respectively.}
		\label{fig:Tc stress 3 components}
\end{center}
\end{figure}

Analogously to the previous section, we fix band filling $n=1.60777$ such that $U(1)$ and $\mathbb{Z}_2$ critical temperatures coincide for a tetragonal system.
We investigate the dependence of $T_c^{U(1)}$ and $T_c^{\mathbb{Z}_2}$ on external stress ($\delta t$) that is shown in \figref{fig:Tc stress 3 components}.
Again, we see four different behaviors for the BTRS critical temperature:
(i) monotonic decrease for large Poisson ratio;
(ii) nonmonotonic behavior for $0.3 \lesssim \nu \lesssim 0.4$;
(iii) fall to zero and re-emergence at larger orthorhombicity for $0.15 \lesssim \nu \lesssim 0.3$;
(iv) steep decrease for small values of Poisson ratio.
Comparison of the regions to the case of only nearest-neighbor interactions ($V_0=0$, \figref{fig:Tc stress 2 components}) shows that values of Poisson ratio at the transitions between different $T_c^{\mathbb{Z}_2}$ behaviors decrease for on-site pairing $V_0 < 0$.
And there is an opposite trend for $V_0 > 0$ (corresponding plots are presented in \appref{app:additional plots}).
Also, the $T_c^{\mathbb{Z}_2}$ re-emergence region shifts towards smaller (larger) external stress values for $V_0 < 0$ ($V_0 > 0$).
Another significant difference is that for $V_0 < 0$ we observe crossing the lines corresponding to different Poisson ratios on $T_c^{\mathbb{Z}_2} (\delta t)$ plot (\figref{fig:Tc stress 3 components}).
There are no such intersections on the $T_c^{\mathbb{Z}_2} (\delta t)$ plot for $V_0 \leq 0$ (see for example \figref{fig:Tc stress 2 components} where $\nu = 0$ and $\nu = 1$ define the boundaries for $\mathbb{Z}_2$ critical temperature).

\begin{figure}
    \begin{center}
		\includegraphics[width=0.8\columnwidth]{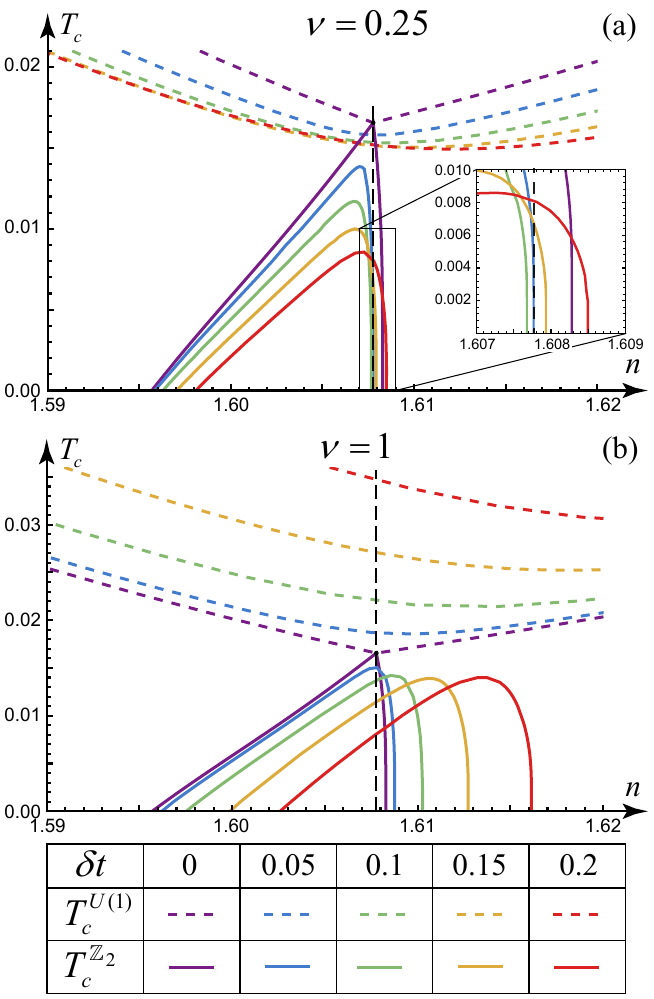}
		\caption{Superconducting phase diagrams in band filling $n$, temperature $T$ coordinates for Poisson ratios
        (a) $\nu=0.25$, (b) $\nu=1$.
        Dashed and solid lines correspond to the $U(1)$ and $\mathbb{Z}_2$ symmetry-breaking critical temperatures, respectively.
        Above the dashed line is a normal metal phase, between the dashed and solid lines is $U(1)$ symmetry-breaking phase, and below the solid line is the BTRS phase.
        BTRS dome first moves towards lower and then towards larger band filling under stress for $\nu=0.25$.
        BTRS dome moves monotonously to a larger band filling under stress for $\nu=1$.
        Vertical black dashed line corresponds to band filling $n=1.60777$ for which $T_c^{U(1)}=T_c^{\mathbb{Z}_2}$ for tetragonal system ($\delta t = 0$).
        On-site and nearest-neighbor interaction strengths are $V_0 = -0.5$ and $V_1=2$, respectively.}
		\label{fig:phase diagrams 3 components}
\end{center}
\end{figure}

We plot $n-T$ phase diagrams for two Poisson ratios (0.25 and 1; \figref{fig:phase diagrams 3 components}) to look at the origin of line intersection in \figref{fig:Tc stress 3 components}.
There is a $s+\sext+d$ phase below the $T_c^{U(1)}$ line for an orthorhombic system.
There are a few differences in the behavior of the BTRS dome for $V_0 = 0$ (\figref{fig:phase diagrams 2 components}) and $V_0 = -0.5$ (\figref{fig:phase diagrams 3 components}) for a strained sample.
(i) BTRS dome for $V_0=-0.5$ shifts $15-20 \%$ more towards larger band filling region than BTRS dome for $V_0=0$ for $\nu=1$.
(ii) Amplitude of the BTRS dome displacement towards smaller band filling (for $\delta t \lesssim 0.05$) and larger band filling (for $\delta t \gtrsim 0.1$) is bigger for $V_0=-0.5$ than for $V_0=0$ for intermediate Poisson ratio values.
Therefore, BTRS dome displacement along the band filling axis for a system with $V_0>0$ ($V_0<0$) is more (less) sensitive to strain than for the $V_0=0$ case.

In this section, we showed that the re-emergent BTRS phase under external stress is a general property for the considered range of interaction constants: On-site pairing $V_0 \in [-0.5;0.5]$ and nearest-neighbor pairing $V_1=2$.
It disappears only for a large positive on-site attraction (see additional plots in \appref{app:additional plots}) in the region $0.5 < V_0 < 1$.
Hence, the complex beyond GL description behavior persists over a certain generalization of the nonlocal interaction model.

\section{Shear and isotropic strain} \label{sec:shear and isotropic}
The behavior of critical temperatures in the vicinity of zero strain and split/unsplit of $T_c^{U(1)}$ and $T_c^{\mathbb{Z}_2}$ under strain are important markers to distinguish different symmetries of superconducting order parameter \cite{grinenko2021unsplit,grinenko2021split,ghosh2025elastocaloric,mattoni2025direct,jerzembeck2024t}.
This section contains the calculation of the response of superconducting and BTRS critical temperatures with respect to shear $\varepsilon_{xy}$ and isotropic (hydrostatic) $\varepsilon_{xx} + \varepsilon_{yy}$ strain.
The Poisson effect does not affect these types of strain.
Hence, we make focus on comparison cases when $T_c^{U(1)}=T_c^{\mathbb{Z}_2}$ and $T_c^{U(1)} > T_c^{\mathbb{Z}_2}$ for zero strain.
This aspect of proximity or identity of two critical temperatures is debated for Sr$_2$RuO$_4$ \cite{grinenko2021split,grinenko2021unsplit}.
\subsection{Shear strain}
Shear strain $\varepsilon_{xy}$ can be implemented in the tight-binding model by adding hoppings $\delta t_{xy}$ with a positive sign along one diagonal and a negative sign for the other diagonal.
The method was proposed in Ref.~\cite{yuan2021strain}.
It leads to dispersion relation
\begin{equation} \label{eq:dispersion relation shear}
    \xi(\bk) = -2 (\cos k_x + \cos k_y) + 4 \delta t_{xy} \sin k_x \sin k_y - \mu.
\end{equation}
There are pure $\sext$-wave, pure $d_{x^2-y^2}$-wave, and $s+id$ state with $\pi/2$ phase difference for a system under shear strain.
We used the symmetry reasoning w.r.t. $k_x \leftrightarrow k_y$ from \secref{sec:nearest neighbor}.
Recall that the implementation of shear $\varepsilon_{xy}$ strain is identical to [100] uniaxial strain.
Hence, the results below are universal for both types of strain.

\begin{figure}[h]
    \begin{center}
		\includegraphics[width=0.8\columnwidth]{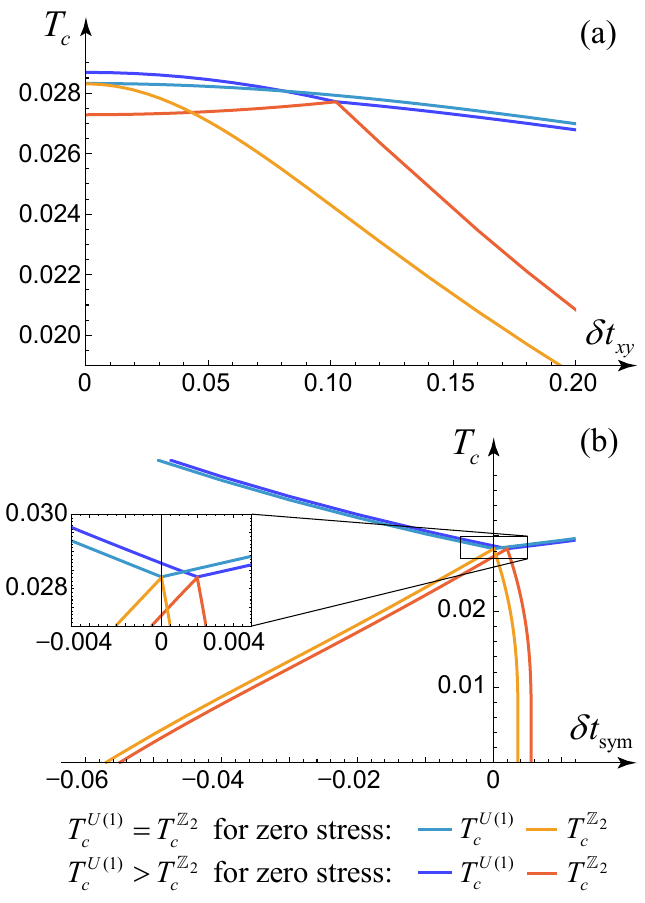}
		\caption{(a) Superconducting and BTRS critical temperatures as a function of shear strain ($\delta t_{xy}$) within microscopic model. 
        Note the absence of a kink in critical temperatures at zero stress for all lines.
        Critical temperatures are symmetric for negative shear strain values.
        (b) Superconducting and BTRS critical temperatures as a function of isotropic strain $\delta t_\text{sym}$ within microscopic model.
        The effect of isotropic strain is similar to a change of the band filling (see purple lines in \figref{fig:phase diagrams 2 components}).
        Case $T_c^{U(1)}>T_c^{\mathbb{Z}_2}$ at zero strain corresponds to band filling $n=1.58458$ and $T_c^{U(1)}=T_c^{\mathbb{Z}_2}$ at zero strain corresponds to $n=1.58518$.}
		\label{fig:microscopic critical temperatures shear and isotropic}
\end{center}
\end{figure}

Figure~\ref{fig:microscopic critical temperatures shear and isotropic}(a) shows shear strain dependence of critical temperatures for two cases: $T_c^{U(1)}>T_c^{\mathbb{Z}_2}$ at zero strain and $T_c^{U(1)}=T_c^{\mathbb{Z}_2}$ at zero strain.
The absence of a kink at zero strain $\varepsilon_{xy}$ can be explained: Shear strain and both gap components belong to different irreducible representations of the tetragonal $D_{4h}$ point group.
Therefore, from the Ginzburg-Landau perspective, shear strain quadratically couples to order parameters.
It excludes the kink in critical temperatures at zero strain that arises from linear coupling between strain and order parameters.

\subsection{Isotropic strain}
Isotropic strain $\varepsilon_{xx} + \varepsilon_{yy}$ can be taken into account by adding third nearest neighbor hoppings $\delta t_\text{sym}$ (as in Ref.~\cite{yuan2021strain})
\begin{equation} \label{eq:dispersion relation isotropic}
    \xi(\bk) = -2 (\cos k_x + \cos k_y) -2 \delta t_\text{sym} (\cos 2k_x + \cos 2k_y) - \mu.
\end{equation}
Isotropic strain preserves the tetragonal symmetry of a lattice.
Hence, possible states are $s$-wave, $d$-wave and $s+id$.

Figure \ref{fig:microscopic critical temperatures shear and isotropic}(b) shows critical temperatures as a function of isotropic stress.
Effect is similar to band filling variation (see purple lines in \figref{fig:phase diagrams 2 components}): $\delta t_\text{sym} > 0$ acts like increase of band filling $n$ in the original model $\xi(\bk) = -2 (\cos k_x + \cos k_y) - \mu$ and vice versa.
Behavior of critical temperatures for the case $T_c^{U(1)}>T_c^{\mathbb{Z}_2}$ at zero strain is identical up to horizontal shift to the behavior of critical temperatures for the case $T_c^{U(1)}=T_c^{\mathbb{Z}_2}$ at zero strain.
The consequence is the presence of a linear kink at nonzero value of isotropic strain for the case $T_c^{U(1)}>T_c^{\mathbb{Z}_2}$ at zero strain.

The general effect of uniaxial, shear, and isotropic strain is a split of superconducting and BTRS critical temperatures in the considered model.
It fits into the picture of accidental degeneracy of the critical temperatures.
Hence, experimental evidence of unsplit (up to error bars) superconducting and BTRS critical temperatures under isotropic strain for Sr$_2$RuO$_4$ \cite{grinenko2021unsplit} points against $s+id$ BTRS superconducting state.

\section{Boundary effects and mesoscopic BTRS states under uniaxial strain} \label{sec:finite system}

\begin{figure}
    \begin{center}
		\includegraphics[width=0.99\linewidth]{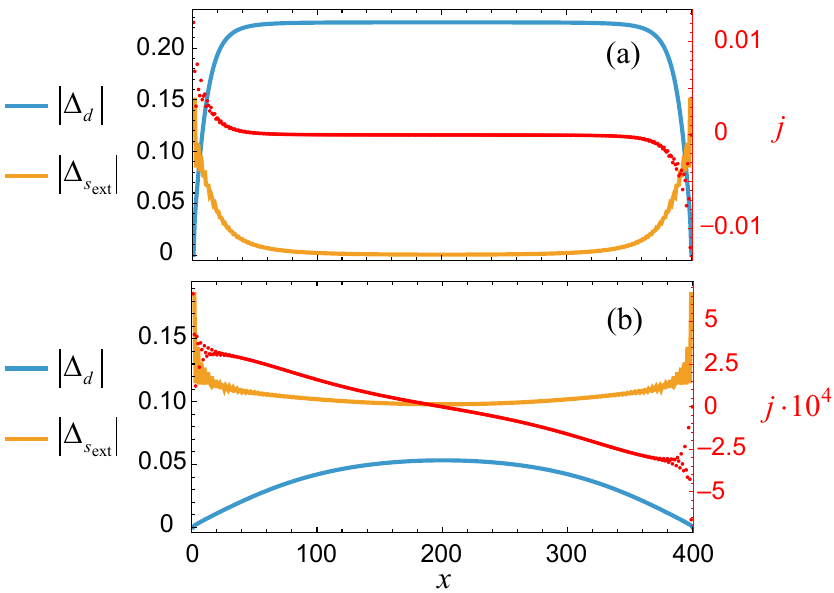}
    \caption{Superconducting gap and current distribution in the cross section of a finite tetragonal sample for (a) boundary BTRS state ($T=0.01$, band filling $n=1.536$), (b) boundary BTRS state that extends over the whole sample ($T=0.01$, $n=1.585$, the gap healing length is comparable or larger than system size).
    Panel (b) is a boundary state because the $d$-wave gap (blue line) does not have a plateau as in panel (a), where bulk equations from \secref{sec:nearest neighbor} apply.
    Phase difference between components is $\pi/2$ in the whole sample.
    Current is in units of $2e/\hbar$.
    Nearest-neighbor interaction strength $V_1=2$.}
     \label{fig:current}
	\end{center}
\end{figure}

To address the existence of superconducting phases with nontrivial broken symmetry, it is also necessary to consider boundary effects, since at boundaries the system can break different symmetries.
Related to this is the question of the stability of BTRS phases in mesoscopic systems.
In this section, we consider a finite sample and show that BTRS states occupy a larger phase diagram region in systems with boundaries than for an infinite sample.
Hence, boundaries provide an additional stabilization mechanism for BTRS states.
This can be explained in the following four-step way.
(i) We know that BTRS states are present when $\sext$-wave and $d$-wave gaps are of the same order (see \figref{fig:gaps and phase difference 2 components});
(ii) $d$-wave order parameter is suppressed near [110] boundary (see \figref{fig:current} and Refs.~\cite{tanuma1999quasiparticle,fogelstrom1997tunneling,honerkamp1999time} for explanations);
(iii) Phase diagram region where bulk $d$-wave amplitude is much greater that $s$-wave amplitude can have 
boundaries with similar amplitudes of gap components.
(iv) It leads to localized boundary BTRS states.
This effect is expected to happen for band filling $n \lesssim 1.6$.
Samples with [100] or [010] boundaries do not host boundary BTRS states since $d$-wave gap is not suppressed next to the boundaries \cite{tanuma1999quasiparticle,fogelstrom1997tunneling,wei1998directional}.

We investigate the band filling--temperature $(n, T)$ phase diagram for a strip-shaped sample.
We consider only nearest-neighbor interaction $V_1=2$ in this section.
Here,  we study a finite $400 \times 400$ system which has [110] surface orientation and periodicity in the transverse direction.
We treat periodic direction using the Fourier transform method applied to Bogoliubov--de Gennes equations \cite{tanuma1999quasiparticle,han2002electronic,han2009method}.
The details of the model, numerical methods, and criteria for different regions in phase diagrams can be found in \appref{app:calculation details}.

\subsection{Stabilization of BTRS state in small systems}

A finite sample with BTRS states has persistent supercurrents (\figref{fig:current}).
If the BTRS phase is localized near the boundaries, supercurrent $j$ is also present only near boundaries [\figref{fig:current}(a)].
The current has a different sign on the opposite edges to satisfy current conservation for the whole system.
If the gap healing length is comparable to the system size, the current extends over the whole sample [\figref{fig:current}(b)].
As a consequence, the gap distribution in \figref{fig:current}(b) does not have plateau in the sample center.
The example of bulk BTRS plateau is \figref{fig:current}(a), where the gap healing length is much smaller than system size.
This effect is specific only to a finite system.

\begin{figure}
    \begin{center}
		\includegraphics[width=0.99\linewidth]{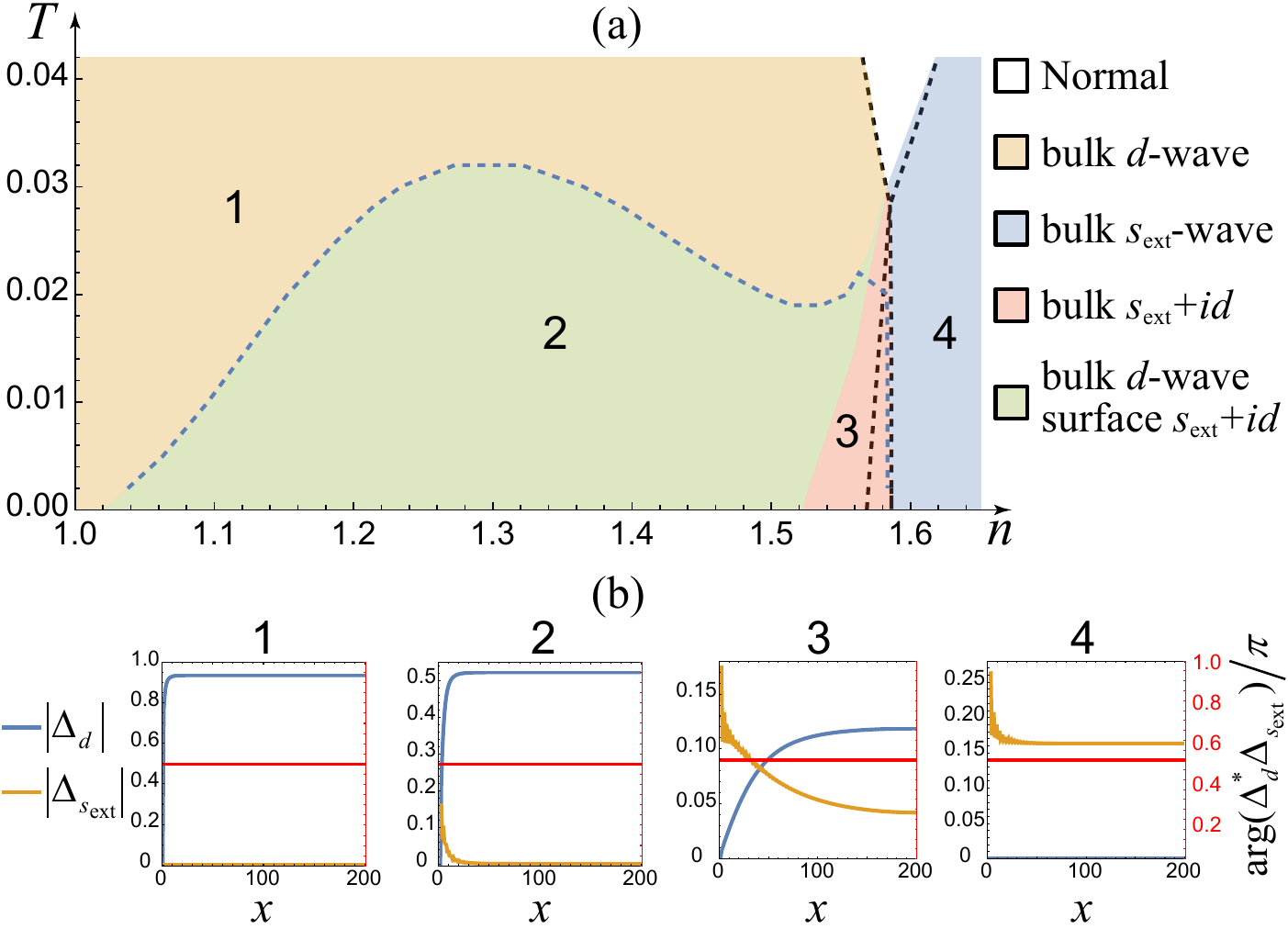}
    \caption{ (a) Superconducting phase diagram in band filling $n$, temperature $T$ coordinates for finite $400 \times 400$ system with [110] surface without strain ($\delta t =0$).
    Differences from infinite system results: (i) BTRS states can be only at the boundaries; (ii) parameter region for BTRS states is wider.
    Large-sized digits enumerate the regions with different gap distributions illustrated in part (b).
    By the bulk properties, we mean properties in the vicinity of the finite sample center.
    Usually, most part of the sample has the same state as in the center.
    Gap components and phase difference are presented for half of the system since the second half is symmetrical.
    Black dashed lines correspond to transition lines for the infinite system (they are identical to purple lines in \figref{fig:phase diagrams 2 components}).
    Blue dashed line envelopes the region with boundary currents in the system ($\max_{\bx} |\boldsymbol{j} (\bx)| > 0.001 \frac{2e}{\hbar}$, like in \figref{fig:current}).
    Nearest-neighbor interaction strength $V_1=2$.
    The phase difference between components is shown to be $\pi/2$ for all phases; however, if one of the gap representations becomes zero (like in  illustrations 1, 2, and 4),
 phase difference becomes ill-defined.}
     \label{fig:finite system no strain phase diagram}
	\end{center}
\end{figure}

Figure~\ref{fig:finite system no strain phase diagram}(a) shows the phase diagram for a finite tetragonal system with [110] surface.
Note the increase in the size of the bulk $\sext + i d$ dome compared to an infinite size system (indicated with black dashed lines).
Localized boundary BTRS states occupy a significant region in $n-T$ parameter space.
Therefore, BTRS states are easier to find in finite mesoscopic samples due to the stabilization mechanism produced by boundaries.
Note that $U(1)$ critical temperature for a finite sample is higher than the critical temperature for the infinite system for band filling $n \gtrsim1.6$.
This happens due to enhanced $\sext$-wave gap near the boundaries [see \figref{fig:finite system no strain phase diagram}(b)4].
It is consistent with previous studies of other $s$-wave models \cite{samoilenka2020boundary,samoilenka2020pair, SamoilenkaPhD,roos2025bcs,hainzl2023boundary}.

\begin{figure}
    \begin{center}
		\includegraphics[width=0.99\linewidth]{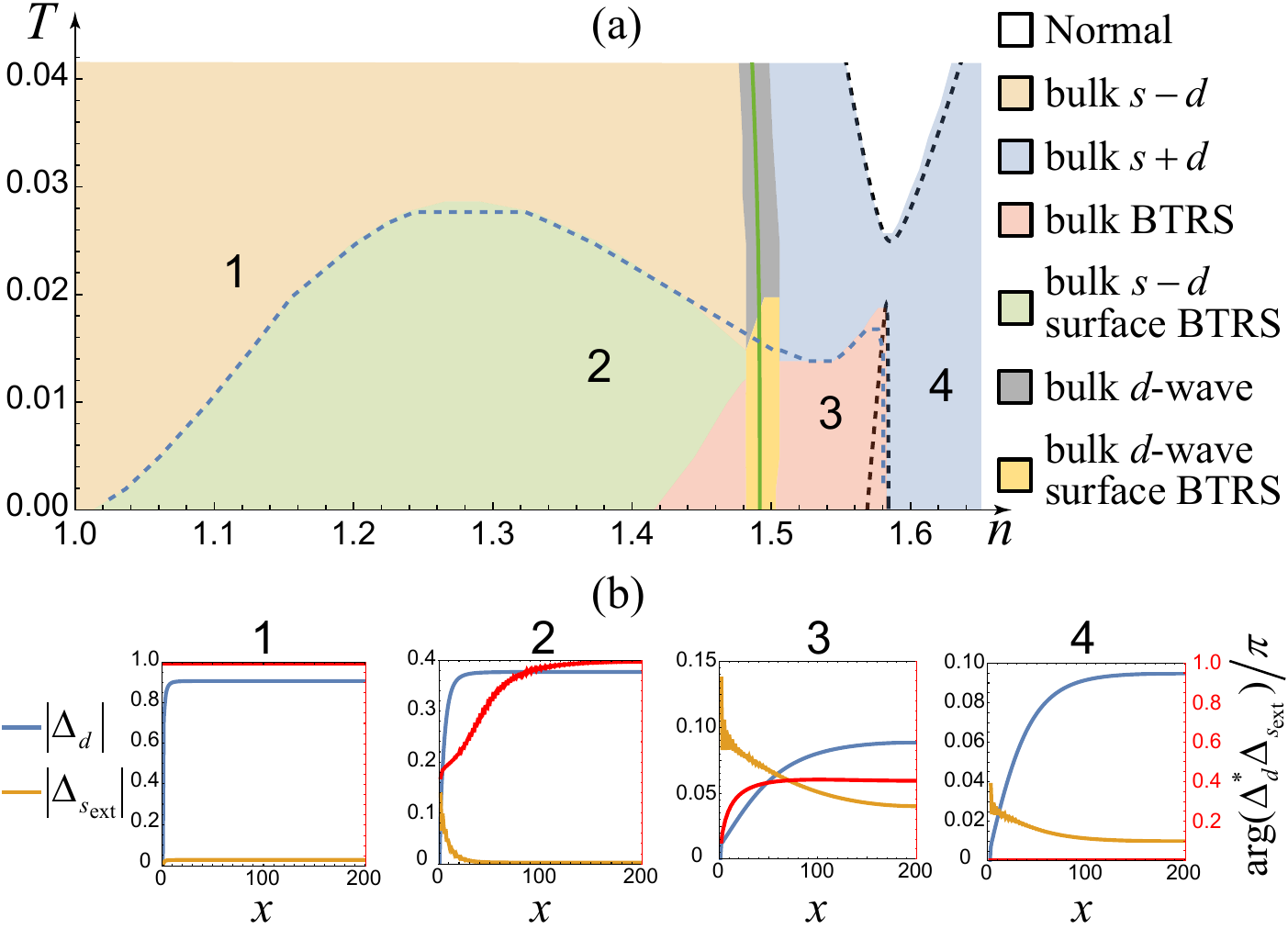}
    \caption{(a) Superconducting phase diagram in band filling $n$, temperature $T$ coordinates for finite $400 \times 400$ system with [110] surface under [100] uniaxial stress.
    Large-sized digits enumerate the regions with different gap distributions illustrated in part (b).
    Gap components (black axis) and phase difference (red axis) are presented for half of a system since the second half is symmetrical.
    Black dashed lines correspond to transition lines for the infinite system [they are identical to green lines in \figref{fig:phase diagrams 2 components}(a)].
    The green line denotes the crossover that separates  $\sext-d$ (to the left) and $\sext+d$ (to the right) phases for the infinite system. The crossover happens by nullifying the s-wave gap (i.e., along this line $\Delta_{\sext}=0$).
    Note the green line indicating the transition between phases $\sext \pm d$ for an infinite system.
    It widens to gray and yellow regions for a finite system (we used the criterion $|\Delta_{\sext}|<0.001$ in the sample center).
    The transition widening is a generic effect for finite system size studies.
    Blue dashed line envelopes the region with boundary currents in the system  ($\max_{\bx} |\boldsymbol{j} (\bx)| > 0.001 \frac{2e}{\hbar}$, like in \figref{fig:current}).
    Nearest-neighbor interaction strength $V_1=2$, measure of orthorhombicity $\delta t =0.1$, Poisson ratio $\nu = 0$.}
     \label{fig:finite system strain 0.1 nu=0 phase diagram}
	\end{center}
\end{figure}

\subsection{Strain effects in small systems}

Next, let us look at the phase diagram for the system under [100] uniaxial stress [$\delta t =0.1, \, \nu = 0$, \figref{fig:finite system strain 0.1 nu=0 phase diagram}(a)].
The general trend is identical to infinite system results from \secref{sec:nearest neighbor}.
The height of the BTRS dome decreases, and the horizontal size (band filling range) increases for the strained system.
The difference is that the BTRS dome is fully surrounded by $\sext+d$ phase for the infinite lattice case (see \secref{sec:nearest neighbor}) and is surrounded by both $\sext \pm d$ phases for the finite system under strain [\figref{fig:finite system strain 0.1 nu=0 phase diagram}(a)].

Note the enhanced $U(1)$ critical temperature for normal metal to $\sext-d$ phase for band filling $n>1.6$ compared to infinite system (black dashed line in \figref{fig:finite system strain 0.1 nu=0 phase diagram}).
The $\sext$ gap component is enhanced near the boundary and exhibits short-scale boundary oscillations.
Since $\sext$ gap is larger than $d$-wave component next to the boundary, it defines the boundary properties of a system.
We know that $s$-wave superconductor has an enhanced boundary critical temperature for intermediate band filling $n$ \cite{samoilenka2020boundary, SamoilenkaPhD}.
Therefore, $\sext-d$ phase can have an enhanced boundary critical temperature for intermediate $n$.

To detect bulk or boundary BTRS state, one can measure the density of states (differential conductance) on the [110] surface.
If a system has $d$-wave or $\sext \pm d$ phase, there will be a zero-bias conductance peak at zero energy \cite{PhysRevLett.72.1526,fogelstrom1997tunneling,tanuma1999quasiparticle,honerkamp1999time}.
If the boundary has $\sext+id$ phase, the peak splits to the two \cite{wei1998directional,tanuma1999quasiparticle,honerkamp1999time,fogelstrom1997tunneling} because the energy gap opens in $\sext$ channel.
Boundary BTRS states produce persistent currents that can also be detected using sensitive magnetometry measurements.

The results of this section show 
that finite samples (especially one or a few coherence lengths size) have richer phase diagrams and new effects.  There are wide parameter regions of BTRS states with persistent currents along boundaries (compare with the different situation  $s+is$ state that can have spontaneous currents in the corners \cite{benfenati2022spontaneous}).
Overall, finite samples can have bulk BTRS states in a much wider band filling region compared to an infinite system.
Hence, boundaries provide a stabilization mechanism for BTRS states, as explained above.

\section{Conclusion} \label{sec:conclusion}

We considered the influence of external stress on a superconductor that hosts $s$-wave and $d$-wave symmetry superconducting components and can break time-reversal symmetry.
In particular, we investigated the effects of the Poisson ratio parameter (describing transverse strain in response to the longitudinal external strain).
The main results are the following:

The microscopic calculations show a significantly more complex phase diagram under [100] uniaxial strain than Ginzburg-Landau-based estimates.

Conventionally, the BTRS phase is dome-shaped on band filling--temperature phase diagram.
The dome touches $U(1)$ critical temperature curve for a tetragonal system.

Including a non-zero Poisson ratio gives rise to the following effects.
The BTRS dome decreases in height (and splits from $T_c^{U(1)}$ curve) under compressional stress for all considered Poisson ratio values.
The dome moves towards a lower (larger) band filling for small (large) values of Poisson ratio.
However, in the parameter space of the model, for intermediate values of Poisson ratio, the BTRS dome first moves to the lower band filling $n$ and then towards larger $n$.
This results in a drop (which can be down to 0) and further increase of $\mathbb{Z}_2$ critical temperature under the increase of compressional stress.
The changing position of the BTRS dome in the phase diagram suggests that there can be materials that have $U(1)$ symmetry breaking state in the tetragonal phase but can exhibit BTRS superconductivity at low temperature under stress.
This can be realized if a tetragonal system is in the proximity of the BTRS phase [band filling of a system is close to the BTRS dome on $(n,T)$ phase diagram].

Another difference with the Ginzburg-Landau model arises in the behavior of the superconducting critical temperature. 
The simplest Ginzburg-Landau approach predicts an increase of $U(1)$ superconducting critical temperature under [100] uniaxial stress \cite{li1993mixed}.
However, our microscopic calculation, including the Poisson effect, shows that the behavior can be significantly different.
Instead, $U(1)$ critical temperature decreases under compressive uniaxial stress in a broad Poisson ratio region but always increases for tensile uniaxial stress.
 Therefore, stress-induced effects under certain conditions become non-monotonous, beyond the validity of the Ginzburg-Landau model.

In the second part of the paper, we explored the effects of boundaries and the finiteness of the sample as a route to control the BTRS state.
In the considered regimes, the existence of boundaries, the finiteness of the samples lead to enhancement of the BTRS state.
A finite system with [110] surface has BTRS boundary states in a wide range of band filling as well as BTRS states extending through the entire mesoscopic sample.
Hence, boundaries provide an additional stabilization mechanism for BTRS states.
A finite system with [100] surface does not have the BTRS boundary states but can host bulk BTRS states.
The BTRS state has persistent superconducting boundary currents for the considered finite systems.
That can be used to detect this form of superconductivity.

Abrupt drop of $\mathbb{Z}_2$ critical temperature for materials with low values of Poisson ratio, can be potentially used in sensitive strain detectors.
For example, if the material is used as a substrate for a sample that exhibits a structural transition, a small strain change leads to a high critical temperature change in the substrate, which can potentially be detected by spontaneous currents generating magnetic fields. This strong dependence of critical temperature on strain also implies it may often be difficult to detect BTRS phase transition in calorimetry probes: the small specific heat feature at BTRS transition can be easily washed out by random strains. 

The second broader implication of these mean-field results concerns systems with strong fluctuations. The strong fluctuations can stabilize the electron quadrupling state: breaking of time-reversal symmetry in the resistive state: \cite{grinenko2021state,bojesen2014phase,bojesen2013time,maccari2023prediction}. The nontrivial change in mean-field critical temperature implies that the size of the electron quadrupling state will also change nontrivially in fluctuating systems.
 
The third aspect is the existence of two tunable critical points in the system. The sensitivity to external perturbations, arising in proximity to a single critical point in ordinary superconductors, was utilized for the creation of single-photon detectors \cite{steinhauer2021progress}. Two tunable critical points in the BTRS system, and their tunability explored here, can be harvested in new detection schemes (see discussion of a different multi-band-superconductors-based scheme \cite{bauer2025type15snspdinteractingvortex}). 

\begin{acknowledgments}

We thank Albert Samoilenka and Vadim Grinenko for useful comments.
This work was supported by the Knut and Alice Wallenberg Foundation via the Wallenberg Center for Quantum Technology (WACQT) and
by the Swedish Research Council Grant 2022-04763. E.B. was supported by Olle Engkvists Stiftelse a project grant from Knut och Alice Wallenbergs Stiftelse,  and partially by the Wallenberg Initiative Materials Science
for Sustainability (WISE) funded by the Knut and Alice Wallenberg Foundation.
The computations were enabled by resources provided by the National Academic Infrastructure for Supercomputing in Sweden (NAISS), partially funded by the Swedish Research Council through Grant Agreement No. 2022-06725.

\end{acknowledgments}

\appendix

\section{Ginzburg-Landau approach for $\sext$-wave and $d$-wave mixing} \label{app:GL method}

Free energy for two component (with $\sext$-wave and $d$-wave symmetries) order parameter in orthorhombic system reads \cite{li1993mixed}
\begin{equation} \label{eq:GL model with strain for s+id}
\begin{split}
    F = &\alpha_1 |\psi_{\sext}|^2 + \alpha_2 |\psi_d|^2 + \beta_1 |\psi_{\sext}|^4 + \beta_2 |\psi_d|^4 \\
    &+  \beta_3 |\psi_{\sext}|^2 |\psi_d|^2+ \frac{\beta_3}{4} (\psi_{\sext}^2 \psi_d^{*2} + \psi_{\sext}^{*2} \psi_d^2) \\
    & + r (\psi_{\sext} \psi_d^{*} + \psi_{\sext}^{*} \psi_d),
\end{split}
\end{equation}
where $r \propto \varepsilon_{xx} - \varepsilon_{yy}$ is a measure of orthorhombicity, $\varepsilon_{xx}\, (\varepsilon_{yy})$ is a strain in the $x \, (y)$ direction.
Coefficients $\alpha_i, \, \beta_i$ can be calculated from microscopic Hamiltonian \eqref{eq:mean-field_Hamiltonian} for tetragonal system \cite{feder1997microscopic}.
Figure~\ref{fig:GL critical temperatures} is computed numerically for the band filling $n=1.58518$ (chemical potential $\mu=1.7976$) and nearest-neighbor interaction $V_1=2$ such that it describes the system from \secref{sec:nearest neighbor}.
In \figref{fig:GL critical temperatures}, system with different $T_c^{U(1)}$ and $T_c^{\mathbb{Z}_2}$ is computed for the band filling $n=1.58458$ (chemical potential $\mu=1.795$).

Superconducting $U(1)$ transition is defined by equating to zero the smallest eigenvalue of the quadratic part of \eqref{eq:GL model with strain for s+id}
\begin{equation}
\begin{split}
    0 &= \text{min} \left(\frac{\alpha_1+ \alpha_2}{2} \pm \sqrt{r^2 + \tfrac{1}{4}(\alpha_1 - \alpha_2)^2} \right) \\
    & \approx
    \begin{cases}
        \text{min}(\alpha_1, \alpha_2) - \frac{r^2}{|\alpha_2 - \alpha_1|}, \, |r| \ll |\alpha_2 - \alpha_1|, \\
        \frac{\alpha_1 + \alpha_2}{2} - |r|, \, |r| \gg |\alpha_2 - \alpha_1|.
    \end{cases}
\end{split}
\end{equation}
Here coefficients can be approximated as $\alpha_i \propto (T - T_{c \, i})$.
Therefore, the normal metal to superconductor transition temperature is higher than $\text{max}(T_{c \, s}, T_{c \, d})$ in GL description.
The corresponding eigenvector of the quadratic part of \eqref{eq:GL model with strain for s+id} defines the order parameter ratio just below the transition.
The eigenvectors are of mixed type with 0 or $\pi$ phase difference $\arg (\psi_s \psi_d^*)$ for nonzero [100] uniaxial strain in the $U(1)$ symmetry breaking phase.

\section{Additional plots for numerical calculations} \label{app:additional plots}

\begin{figure}[h]
    \begin{center}
		\includegraphics[width=0.85\columnwidth]{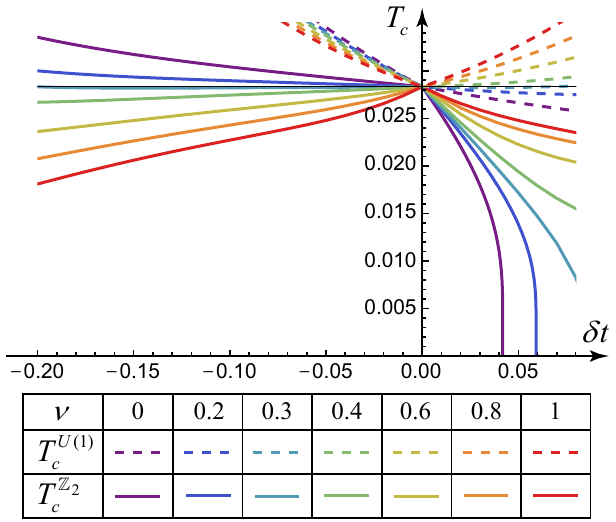}
		\caption{Superconducting and BTRS critical temperatures as a function of change of hopping $\delta t$ (a measure of orthorhombicity) due to [100] uniaxial stress for different Poisson ratios $\nu$.
        Positive (negative) values of $\delta t$ correspond to compressional (tensile) stress in the [100] direction.
        Compressive and tensile stress are not symmetric for $\nu \neq 1$.
        Note linear kink for both critical temperatures at zero strain, consistent with GL theory.
        The right part of the plot ($\delta t > 0$) is identical to \figref{fig:Tc stress 2 components}.
        Dashed and solid lines correspond to the $U(1)$ and $\mathbb{Z}_2$ symmetry-breaking critical temperatures, respectively.
        The horizontal black line indicates critical temperature for the tetragonal system without stress $T_c^{U(1)}=T_c^{\mathbb{Z}_2}=0.02832$, $\delta t =0$, band filling $n=1.58518$.
        Nearest-neighbor interaction strength $V_1=2$.}
		\label{fig:Tc tensile stress 2 components additional}
\end{center}
\end{figure}

First, we present plot of $T_c^{U(1)}$ and $T_c^{\mathbb{Z}_2}$ for nearest-neighbor interaction potential (like in \secref{sec:nearest neighbor}) as a function of \textit{tensile} [100] uniaxial stress ($\delta t < 0$) illustrated in \figref{fig:Tc tensile stress 2 components additional}.
Obviously, the system has symmetry w.r.t. change $\delta t \rightarrow \nu' \delta t', \, \nu \rightarrow 1/\nu', \, \Delta_{\sext} \rightarrow - \Delta_{\sext}'$, where ($\delta t' < 0$ corresponds to a compressional stress $|\delta t'|$ in the [010] direction).
The sign change of one of the order parameter components comes from the linear dependence of the off-diagonal matrix element on strain ($S_{d, \, \sext} \propto \delta t$) in the lowest order expansion.
Therefore, we expect to have both $T_c^{U(1)}$ and $T_c^{\mathbb{Z}_2}$ order for different $\nu$ to be opposite to the one presented in \figref{fig:Tc stress 2 components} (we predict the following order: The larger $\nu$ the lower $T_c^{\mathbb{Z}_2}$ and $T_c^{U(1)}$).
Another note from the symmetry is that for a Poisson ratio equal to unity, positive and negative strain lead to identical effects in both critical temperatures (like GL calculations in \figref{fig:GL critical temperatures}).
We observe monotonous dependencies of critical temperature as a function of tensile stress for all Poisson ratio values.
Note unusual behavior for $\nu=0.3$: $\mathbb{Z}_2$ critical temperature remains approximately constant for tensile stress $\delta t \in [-0.2; 0]$.

\begin{figure}[h]
    \begin{center}
		\includegraphics[width=0.85\columnwidth]{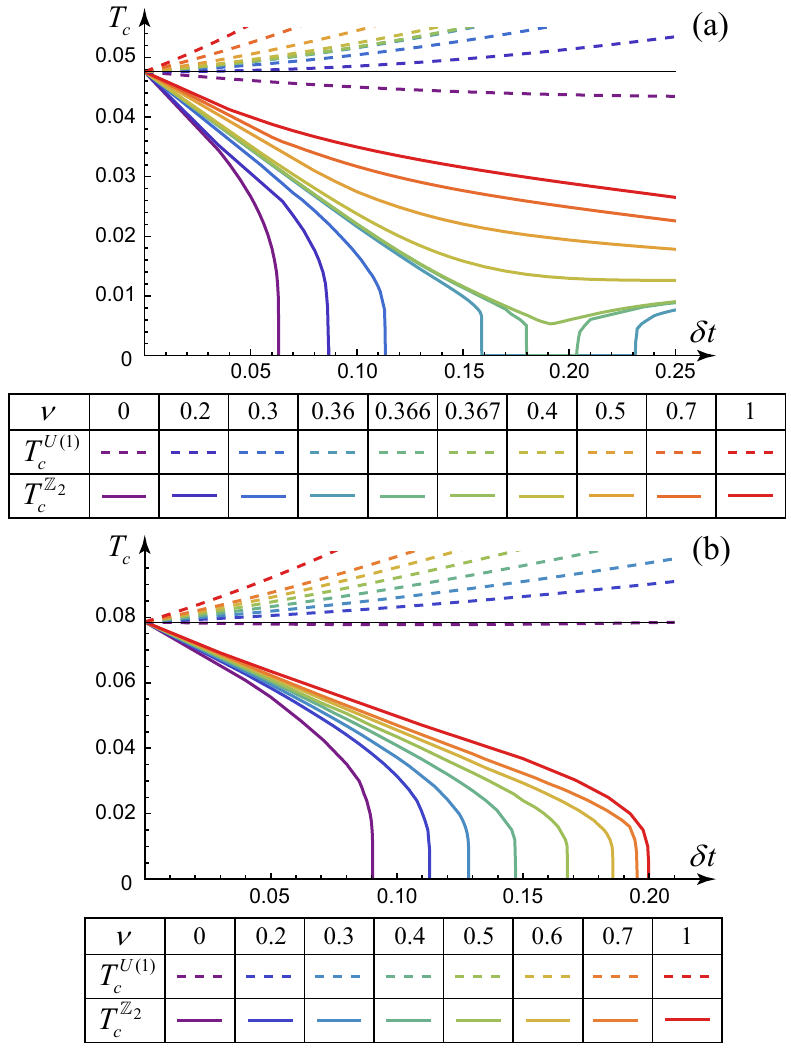}
		\caption{Superconducting and BTRS critical temperatures as a function of change of hopping $\delta t$ (a measure of orthorhombicity) due to [100] uniaxial compressional stress for different Poisson ratios $\nu$.
        (a) On-site interaction strength $V_0=0.5$, (b) $V_0=1$.
        Nearest-neighbor interaction strength $V_1=2$.
        Dashed and solid lines correspond to the $U(1)$ and $\mathbb{Z}_2$ symmetry-breaking critical temperatures, respectively.
        Increasing on-site attraction pairing leads to the shift of an abrupt $\mathbb{Z}_2$ critical temperature drop towards larger stress values.
        The horizontal black line indicates critical temperature for the tetragonal system without stress $T_c^{U(1)}=T_c^{\mathbb{Z}_2}=0.04759$, $\delta t =0$, band filling $n=1.55855$ for (a) and $T_c^{U(1)}=T_c^{\mathbb{Z}_2}=0.07859$, $n=1.52612$ for (b). Nearest-neighbor interaction strength $V_1=2$.}
		\label{fig:Tc stress 3 components additional}
\end{center}
\end{figure}

In the section \ref{sec:additional onsite interaction} and in the Appendix, we avoid studying large negative values of on-site interaction $V_0$ (corresponding to on-site repulsion) to prevent competition with magnetic order parameter in the system.
Therefore, we present additional plots for positive values of the $V_0$.
Figure~\ref{fig:Tc stress 3 components additional} shows the dependence (split) of $T_c^{U(1)}$ and $T_c^{\mathbb{Z}_2}$ on [100] uniaxial strain ($\delta t$) by analogy with Sections \ref{sec:nearest neighbor} and \ref{sec:additional onsite interaction}.
We observe drop to zero and re-emergence of $T_c^{\mathbb{Z}_2}$ for $V_0 = 0.5$ and Poisson ratio $\nu \in [0.36;0.366]$  [\figref{fig:Tc stress 3 components additional}(a)].
However, the $T_c^{\mathbb{Z}_2} (\delta t)$ has trivial monotonic drop to zero for $V_0 = 1$ [\figref{fig:Tc stress 3 components additional}(b)].

\begin{figure}[h]
    \begin{center}
		\includegraphics[width=0.85\columnwidth]{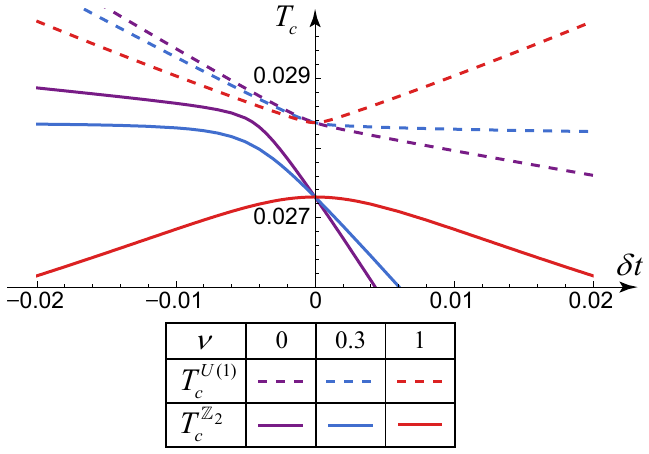}
		\caption{Superconducting and BTRS critical temperatures as a function of tensile ($\delta t <0$) and compressional ($\delta t >0$) [100] uniaxial strain within microscopic model. 
        System has $T_c^{U(1)}>T_c^{\mathbb{Z}_2}$ at zero external strain, $n=1.58530$.
        Note the absence of kink in critical temperatures at zero stress.
        Compressive and tensile stress are not symmetric for $\nu \neq 1$.}
		\label{fig:microscopic critical temperatures small stress n=1.5853}
\end{center}
\end{figure}

\begin{figure}[h]
    \begin{center}
		\includegraphics[width=0.9\columnwidth]{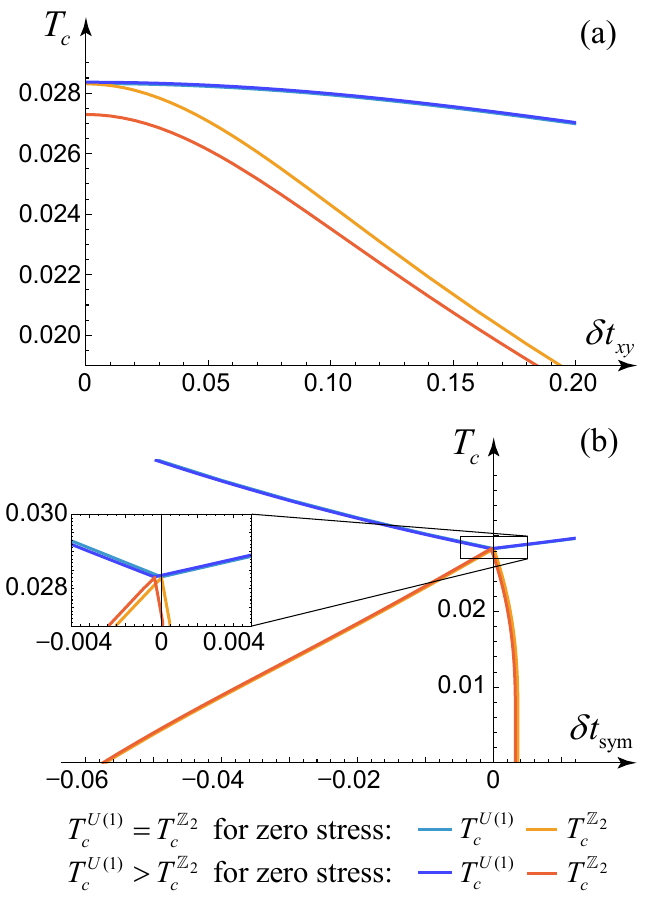}
		\caption{(a) Superconducting and BTRS critical temperatures as a function of shear strain ($\delta t_{xy}$) within microscopic model. 
        Note the absence of kink in critical temperatures at zero stress for all lines.
        Critical temperatures are symmetric for negative shear strain values.
        (b) Superconducting and BTRS critical temperatures as a function of isotropic strain $\delta t_\text{sym}$ within microscopic model.
        Case $T_c^{U(1)}>T_c^{\mathbb{Z}_2}$ at zero strain corresponds to band filling $n=1.58530$ and $T_c^{U(1)}=T_c^{\mathbb{Z}_2}$ at zero strain corresponds to $n=1.58518$.}
		\label{fig:microscopic critical temperatures shear and isotropic n=1.5853}
\end{center}
\end{figure}

In sections \ref{sec:nearest neighbor} and \ref{sec:shear and isotropic}, we consider a case when superconducting and BTRS critical temperatures are split at zero stress.
We choose band filling $n=1.58458$ to illustrate results in Figs.~\ref{fig:microscopic critical temperatures small stress}, \ref{fig:microscopic critical temperatures shear and isotropic}.
This filling is smaller than "optimal" one: $n=1.58518$ that corresponds to $T_c^{U(1)}=T_c^{\mathbb{Z}_2}$ at zero strain.
In this case, $U(1)$ symmetry breaking phase corresponds to pure $d$-wave at zero stress.
In order to address question: What changes if one takes band filling larger than "optimal" one?
We choose band filling $n=1.58530$ that corresponds to  $T_c^{U(1)}>T_c^{\mathbb{Z}_2}$ situation (see intersection of purple lines with $n=1.5853$ in \figref{fig:phase diagrams 2 components}).
In this case, $U(1)$ symmetry breaking phase corresponds to pure $s$-wave at zero stress.
Figures \ref{fig:microscopic critical temperatures small stress n=1.5853} and \ref{fig:microscopic critical temperatures shear and isotropic n=1.5853} present calculations of critical temperatures response to uniaxial, shear, and isotropic strain for $n=1.5853$.

\section{Finite system calculation details} \label{app:calculation details}
Discrete analog of the model in \secref{sec:general model} corresponds to the following mean-field Hamiltonian
\begin{equation}\label{eq:mean-field_Hamiltonian}
\begin{split}
    H_\text{MF} = & -\sum_{\bx,\bx', \sigma} h (\bx,\bx') c_{\bx,\sigma}\da c_{\bx',\sigma} \\
    & + \sum_{\langle \bx,\bx'\rangle} \left( \Delta_{\bx,\bx'} c_{\bx,\uparrow}\da c_{\bx',\downarrow}\da + \Delta^{*}_{\bx,\bx'} c_{\bx',\downarrow} c_{\bx,\uparrow}\right) + \text{const},
\end{split}
\end{equation}
where position $\bx=(i,j)$, kinetic energy $h (\bx,\bx') = \mu \delta_{i,i'} \delta_{j,j'} + t_x \delta_{i,i'\pm1} \delta_{j,j'} + t_y \delta_{i,i'} \delta_{j,j'\pm1}$ with hopping integrals $t_x = 1 + \delta t$, $t_y=1-\nu \delta t$, and superconducting order parameter lives on a nearest-neighbor links $\Delta_{\bx,\bx'} = V_1 \langle c_{\bx,\downarrow} c_{\bx',\uparrow} \rangle$.
Here $c_{\bx,\sigma}\da (c_{\bx,\sigma})$ is the creation (annihilation) operator for an electron with spin $\sigma$ on-site $\bx$.
The relation between link order parameter and gap irreducible representations (that are defined on sites) is
\begin{equation}
\begin{split} 
    \Delta_d (\bx) = \tfrac{1}{2}  & \left( \Delta_{\bx,(i+1,j)} + \Delta_{\bx,(i-1,j)} \right. \\
    & \left. - \Delta_{\bx,(i,j+1)} - \Delta_{\bx,(i,j-1)} \right), \\
    \Delta_{\sext} (\bx) = \tfrac{1}{2} & \left( \Delta_{\bx,(i+1,j)} + \Delta_{\bx,(i-1,j)} \right. \\
    & \left. + \Delta_{\bx,(i,j+1)} + \Delta_{\bx,(i,j-1)} \right) .
\end{split}
\end{equation}
On the end sites ($x=0, \, N_x$), the site gap is ill-defined due to the absence of two links.
Therefore we use $\Delta_{d, \, \sext} = 0$ for them.
The averaged current from site $\bx'$ to $\bx$ is given by \cite{zhu2016bogoliubov}
\begin{equation}
    j_{\bx,\bx'} = \frac{2e}{i\hbar} \sum_{\sigma}{\left( t_{\bx,\bx'} \langle c_{\bx,\sigma}\da c_{\bx',\sigma} \rangle - t_{\bx',\bx} \langle c_{\bx',\sigma}\da c_{\bx,\sigma} \rangle \right)}.
\end{equation}

\begin{figure}
    \begin{center}
		\includegraphics[width=0.7\columnwidth]{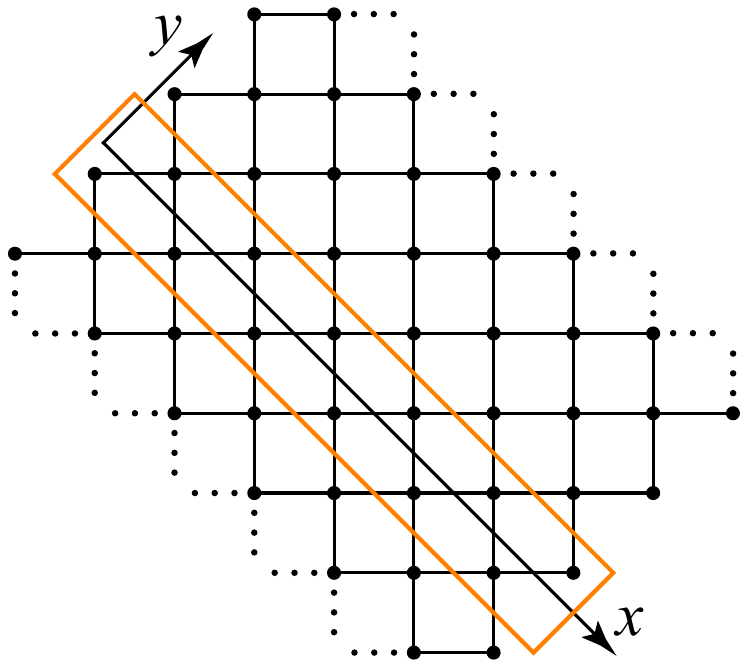}
		\caption{Periodic in the $y$ direction rectangular lattice with [110] surface (using standard axes orientation), which models an infinite strip.
        The orange rectangle shows the supercell for Fourier transform in the $y$ direction and numerical calculations.}
		\label{fig:lattice}
\end{center}
\end{figure}

The system we model is illustrated in \figref{fig:lattice}.
Fourier transform of the supercell (indicated with orange rectangle in \figref{fig:lattice}) in the $y$ direction allows to reduce computational complexity from $\mathcal{O} \left( N_x^3 N_y^3\right)$ to $\mathcal{O} \left( N_x^3 N_y\right)$.
The third power comes from the computation of eigenvalues and eigenvectors of the $2N_xN_y \times 2N_xN_y$ matrix.
The convergence criterion is set to $\max_{\langle \bx,\bx'\rangle} |1 - \Delta_{\bx,\bx'} (m+1) /\Delta_{\bx,\bx'} (m) | < 10^{-7}$, where $m$ is an iteration number.

We use the following criteria for gap components to define the region:
$|\Delta_d (\bx)| > 0.001$, $|\Delta_{\sext}|<0.001$ in the sample center, $\max_{\bx} |\Delta_{\sext} (\bx)| > 0.01$, and phase difference between components is nontrivial near the boundary.
We also compute currents in the system.
Another criterion for BTRS states (bulk or boundary ones) is $\max_{\bx} |\boldsymbol{j} (\bx)| > 0.001 \frac{2e}{\hbar}$ [illustrated with a blue dashed line in \figref{fig:finite system no strain phase diagram}(a)].
This method fails to produce consistent results when both order parameter components are small [region $\{n \in [1.56;1.58], \, T \in [0.022;0.03] \}$ in \figref{fig:finite system no strain phase diagram}(a)].
Bulk $\sext+id$ phase is defined as $|\Delta_{d, \, \sext}| > 0.001$ in the sample center with nontrivial phase difference.

\bibliography{references}

\end{document}